\documentclass[12pt , a4paper]{report}
\usepackage[top=3cm,right=3.5cm,bottom=3cm,left=2.5cm]{geometry}
\usepackage[dvipsnames]{xcolor}
\usepackage{hyperref}
\hypersetup{colorlinks=true, 
	linkcolor= red ,
	filecolor=blue,
	citecolor =green,      
	urlcolor=blue,
}
\usepackage{cite}
\usepackage{tikz}
\usepackage{tikz-3dplot}
\usepackage{verbatim}
\usepackage[nottoc]{tocbibind}
\usepackage{afterpage}
\usepackage{fancyhdr}
\usepackage{amsmath}
\usepackage{amssymb}
\usepackage{textcomp}
\usepackage{subcaption}
\usepackage{float}
\usepackage[hang]{footmisc}
\usepackage{setspace}
\usepackage{indentfirst}
\usepackage{graphicx}
\usepackage{textcomp}
\usepackage{makeidx}
\usepackage{booktabs}
\usepackage{cite}
\usepackage{tikz}
\usepackage{tikz-3dplot}
\usepackage[toc,page]{appendix}
\usepackage[font=small,labelfont=bf]{caption}
\usetikzlibrary{shapes}
\usetikzlibrary{snakes}
\newcommand\blankpage{%
	\null
	\thispagestyle{empty}%
	\addtocounter{page}{-1}%
	\newpage}
\usetikzlibrary{svg.path}
\newcommand{\KN}{\mathbin{\bigcirc\mspace{-15mu}\wedge\mspace{3mu}}}
\begin{document}
		\begin{titlepage}
		\centering
		\includegraphics[width=0.15\textwidth]{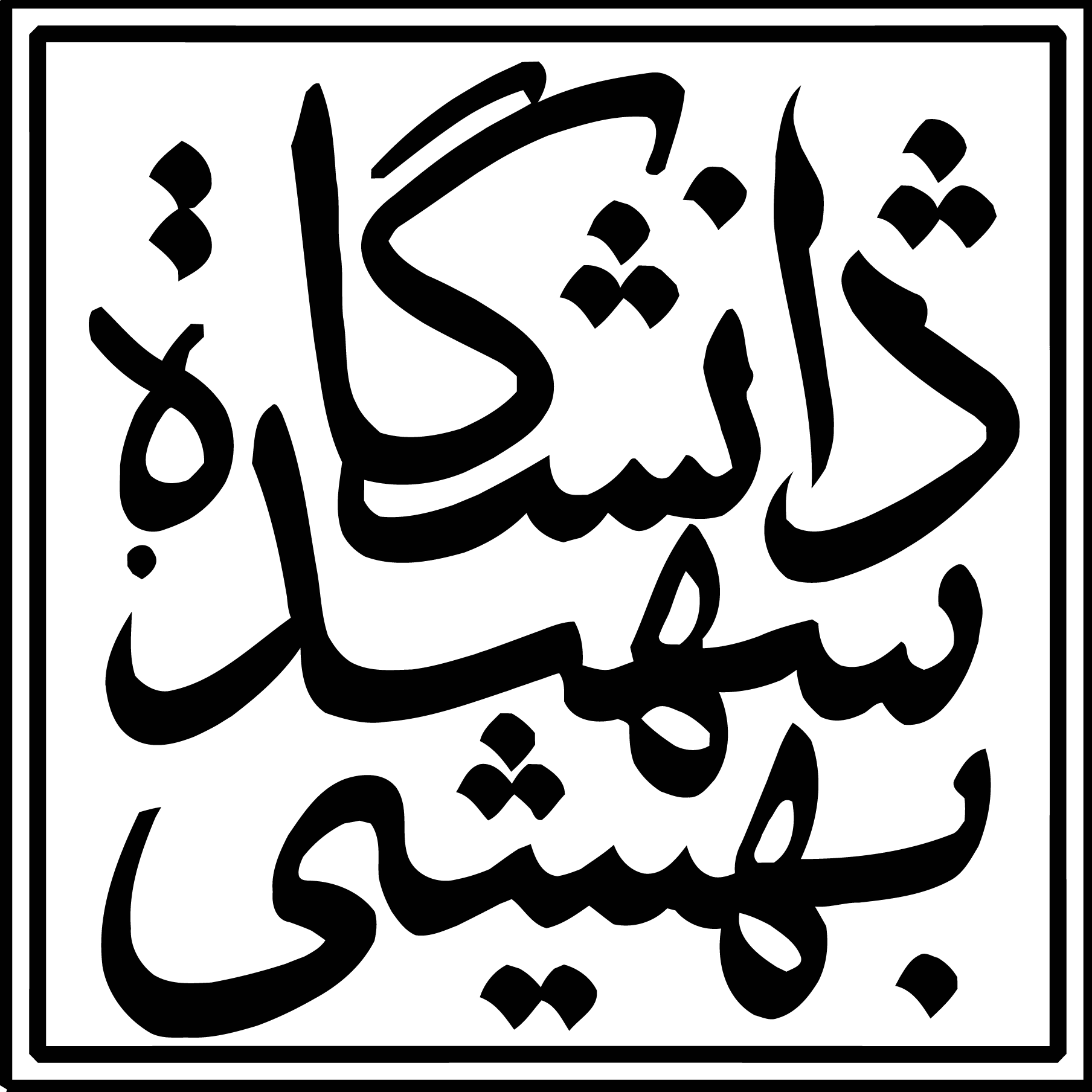}\par
		{\scshape\Large Shahid Beheshti University \par}
		{\scshape\Large Faculty of Physics\par }
		\vspace{1cm}
		{\scshape A DISSERTATION SUBMITTED IN PARTIAL FULFILLMENT OF THE REQUIREMENTS FOR THE DEGREE OF MASTER OF PHYSICS\par}
		\vspace{1.5cm}
		{\huge\bfseries Study of the de Sitter space-time and its behavior at infinity \par}
		\vspace{2cm}
		{\scshape\large By\par}
		{\Large\itshape  Djeyrane-Sophie Erfani Harami\par}
		\vspace{0.3cm}
		{\scshape\large Supervisor\par}
		{\Large\itshape  Dr. Hossein Shojaie\par}
		\vspace{1.5cm}
		{\scshape January 2020\par}
	\end{titlepage}	
\blankpage
\begin{titlepage}
	\centering
	\vspace{15cm}
		{\Large\itshape  In memory of my grandfather!\par}
\end{titlepage}
\blankpage

\begin{abstract}
	The aim of this manuscript is to review the studies about de Sitter solution and the null infinity of asymptotically flat and de Sitter space-times. Thus, after introducing the de Sitter space-time, the symmetry group is described. Also precise definitions of asymptotically flat  and  de Sitter space-times are reviewed. Henceforth,  the null infinities and the asymptotic symmetry groups of these two space-times are considered, which lead to the  Fefferman-Graham approach and the Penrose-Newman formalism. 
	

\end{abstract}
	\tableofcontents
\listoffigures
	\chapter{Introduction}
	Efforts to find the speed of light began in the seventeenth century.
	      It was only then that the first evidence that light had a limited and measurable speed, came to the mind. Until then, it was almost universally believed that the speed of light has an infinite value. Afterward many savants worked on that subject to find the value of the speed of light as Armand Fizeau, Albert Michelson and many others. In  1983 International Bureau of Weights and Measures reported the speed of light to have the value of $299,792,458m/s$.
	
	In 1905 Einstein declared the theory of special relativity based on two postulates. The First one is that  physical laws are invariant in inertial frames. The second one is the invariance of the speed of light. Based only on these two hypotheses and without referring to mechanics laws, electromagnetic and other fundamental physical theories, he derived Lorentz transformations. Keeping the speed of light invariant, these transformations took the place of Galilean transformations.
	
	Another fundamental physical constant is Planck length, $l_p$, which has the length dimension and has been obtained by combining the gravitational constant, $G$, the speed of light, $c$ and the Planck constant $\hbar$ \cite{gaarder2016gravitational}. After finding the constant of action, known as Planck constant, Max Planck mentioned that by using the three constants $G$,
	$c$
and
	$\hbar$, it is possible to define a global constant for length. In General relativity, the distance between two points is a dynamical parameter and is obtained by solving Einstein's field equations. Through the quantum mechanics laws, each dynamical parameter must satisfy the uncertainty principle. Actually, $l_p$ is the distance where quantum effects appear. Lorentz transformations do not preserve this minimum length. As Lorentz transformations replace Galilean transformations, it is expected that a new symmetry group that preserves $l_P$  and $c$, becomes practical.
	
	Lorentz transformations can also be performed in de Sitter space-time as it is homogeneous\cite{aldrovandi2007sitter}. Hence  Minkowski background in physical theories may be replaced by de Sitter space-time that its governing symmetry group is So(4,1) \cite{abbott1982stability,hawking1973large}. These transformations can preserve the length. So it is possible to escape the mentioned problem \cite{aldrovandi2007sitter,cacciatori2008special,nakahara2018geometry}.
	
  	 Also observations show that our universe expands at an accelerating rate and  a model with a positive constant is more appropriate to describe it \cite{riess1998observational}.  Hence it is useful to study de Sitter space-time.  Different topics have to be considered to be sure about this choice. Actually de Sitter horizon complicates many things. As in the flat case background it is important to consider space-time's boundaries \cite{wald1999gravitational,wald2000general}. In presence of a positive cosmological constant,
	de Sitter null infinity, $\mathcal{I}$ is no longer null but spacelike \ and can not be study like asymptotically flat $\mathcal{I}$ which have been study by Bondi et al. in 1962 \cite{bondi1962gravitational}. They rewritten Minkowski metric in Bondi coordinates, $(u,r,x^A)$ where $u=t-r$ is the retarded time and $x^A=(\theta,\varphi)$. They used Dirichlet boundary conditions  and found a meaningful notion of energy. But it is not possible to follow the same route in asymptotically de Sitter space-time because using  these boundary conditions gravitational waves do not carry away de Sitter charges across future null infinity \cite{ashtekar2014asymptotics,ashtekar2015asymptotics}. So the Fefferman-Graham method \cite{fefferman1985conformal,fefferman2011ambient} and Penrose-Newaman formalism \cite{penrose1965remarkable,penrose1984spinors}  can be used to study $\mathcal{I}$ in de Sitter space-time \cite{saw2016mass}. Also it is possible to add  
	Neumann boundary conditions $J^{AB}=0$ to Dirichlet boundary conditions and find finite, conserved, integrable and generically non-vanishing \cite{compere2019lambda}.
	   
	 The de Sitter solution for Einstein's field equations is obtained in chapter two. Also the different coordinate systems for de Sitter space-time are considered. The symmetry group of the de Sitter space-time is obtained in chapter three and it is explained how these transformations can preserve a minimum length. Asymptotically flat \cite{bondi1962gravitational,penrose1965remarkable,newman1966note,arnowitt2008republication,moreschi1987general,bros2002asymptotic} and de Sitter space-times \cite{addazi2020conformal,anderson2005structure, aneesh2019conserved,anninos2011asymptotic,anninos2019sitter,ashtekar2014asymptotics,ashtekar2015asymptotics,ashtekar2019asymptotics,ashtekar2014geometry} are considered in chapter four.
 
		\chapter{De Sitter space-time}
	There exist three maximally symmetric vacuum solutions for	Einstein's equations known  as de Sitter, anti-de Sitter and Minkowski with positive, negative and zero curvature. Here the focus is on Einstein's equations' solution in presence of a positive cosmological constant, then  different coordinate systems and Killing's vector fields are considered.
	\section{De Sitter metric}
	A structure  for symmetric tensor type
	$(0,2)$,
	 known as the metric tensor, has been introduced that relates two vectors in the vector space $T_{p}$ \cite{stephani2009exact, penrose1984spinors}
	 \begin{equation}
	 \eta_{\alpha\beta}e_{\mu}^{\alpha} e_{\nu}^{\alpha}=g_{\mu \nu}
	 \end{equation}
	 where
	 $ \begin{Bmatrix}
	 e_{\mu} ^{\alpha}
	 \end{Bmatrix}$
	 is a null tetrad that consists of two real null vectors
	 $l$,
	 $k$ 
	 and two complex conjugate null vectors 
	 $m$,
	 $\bar{m}$.\cite{penrose1984spinors}
	 \begin{equation}
	 \begin{Bmatrix}
	 e_{\mu}^{\alpha}
	 \end{Bmatrix}
	 =(m,\bar{m},l,k).
	 \end{equation}
	 One can write the length element
	  $ds^{2}$
	 using vector basis
	 in 
	 $T^*_{p}$ as \cite{fukuyama2009comments}
	 \begin{equation}
	 \label{eq:metric}
	 ds^{2}=g_{\mu \nu}\omega^{\mu}\omega^{\nu}.
	 \end{equation}
	 Also by taking  coordinate basis or holonomic frame to the account, equation 
	 \eqref{eq:metric}
	 can turn to 
	 \begin{equation}
	 ds^{2}=g_{\mu \nu}dx^{\mu}dx^{\nu}.
	 \end{equation}
	 Actually,  one can define four spacelike vectors,
	 $e^{\alpha}$ 
	 and a timelike vector,
	 $e^{0}\equiv X^{0}$, in five dimensions
	 as
	 \begin{equation}
	 e^{\mu}
	 =
	 (X^{0},e^{\alpha})
	 =
	 (X^{0},X^{1},X^{2},X^{3},X^{4})
	 .
	 \end{equation}
	  Furthermore relations as follows can be described
	 \begin{equation}
	 e^{\alpha} e_{\beta}=\delta_{\alpha}^{ \beta}\: \: \:,\: \: \: X^{0} X_{0}=-1 \: \: \:,\: \: \: e^{\alpha} X_{0}=0.
	 \end{equation}
	 Then one can write \cite{tod2015some}
	 \begin{equation}
	 l^{2}=-X^{0} X_{0}+X^{1} X_{1}+X^{2} X_{2}+X^{3} X_{3}+X^{4} X_{4}.
	 \label{eq:2.00}
	 \end{equation}
	 Equation \eqref{eq:2.00} shows a hyperbloyd embedded in five dimensional Minkowski space-time with the line element 
	 \begin{equation}
	 ds^2=-dx^{2}_{0}+dx^{2}_{1}+dx^{2}_{2}+dx^{2}_{3}+dx^{2}_{4}.
	 \label{eq:2.200}
	 \end{equation}
	 This relation can also be obtained by solving Einstein's equations (see section \ref{sec:2.2.2}).
	 \section{Solving Einstein's equations}
	 Most physical theories are introduced by mathematical models and are described by a set of differential equations. Among gravitational theories, Einstein's theory has been accepted as the most successful. 
	 In this case, differential equations are written, considering that space and time can be introduced with a pseudo Riemannian manifold and a distribution of interaction of matter and gravity.
	 Usually we search for exact solutions or, if possible, a general solution of differential equations. Many of these exact solutions are not physical but many others like Schwarzschild and Kerr solutions for black holes, Friedmann-Lemaître-Robertson-Walker solution for cosmology are physical \cite{stephani2009exact}. Unless imposing any restrictions on the energy-momentum tensor, each metric can be the solution of these equations, because it is just a description of the 	$T_{\mu \nu}$. We can apply symmetry conditions to the metric, for example by imposing algebraic constraints on the Riemann tensor or selecting boundary conditions. Here are the field equations in presence of cosmological constant
	 	\begin{equation}
	 G_{\mu \nu}+\Lambda g_{\mu \nu}=(8 \pi G /c^{4})T_{\mu \nu},
	 \label{eq:2.1}
	 \end{equation}
	  one can consider the speed of light to be equal to one in relation \eqref{eq:2.1}.
	  
	  Considering the matter field to be zero
	  	$(T_{\mu \nu}=0)$
	   is one of these conditions that simplify equations 	\eqref{eq:2.1}. Solving Einstein equations in presence of cosmological constant for an isotropic homogeneous model, one obtains solutions that will be the simplest inflationary solutions
	   	\begin{equation}
	  R_{\mu \nu} -1/2 R g_{\mu \nu}= -\Lambda g_{\mu \nu}.
	   \label{eq:2.3}
	   \end{equation}
	   Ricci scalar and Ricci tensor for this model are 
	   $R=4 \Lambda$, $R_{\mu \nu}=\Lambda g_{\mu \nu}$.
	 \subsection{Static and spherically symmetric coordinates}
	 To achieve the ability to solve equations 
	 \eqref{eq:2.1}
	 one have to write a basic form of the metric \cite{lenk2010general,gron2007homogeneous,carroll2019spacetime}
	 \begin{equation}
	 g_{\mu \nu}=
	 \begin{bmatrix}
	 A(t,r,\theta,\varphi)&B(t,r,\theta,\varphi)&C(t,r,\theta,\varphi)&D(t,r,\theta,\varphi)\\
	 B(t,r,\theta,\varphi)&E(t,r,\theta,\varphi)&F(t,r,\theta,\varphi)&G(t,r,\theta,\varphi)\\
	 C(t,r,\theta,\varphi)&F(t,r,\theta,\varphi)&H(t,r,\theta,\varphi)&I(t,r,\theta,\varphi)\\
	 D(t,r,\theta,\varphi)&G(t,r,\theta,\varphi)&I(t,r,\theta,\varphi)&J(t,r,\theta,\varphi)
	 \end{bmatrix}.
	 \label{eq:2.5}
	 \end{equation}
	  Ricci tensor can be obtained according to the metric \eqref{eq:2.5} then the result can be used in \eqref{eq:2.1} to find the final form of the metric. Accurately to obtain an exact solution, the metric is considered  to be stationary, which means a timelike Killing vector field exists and a timelike coordinate can be defined according to this Killing vector field
	($\frac{\partial g_{\mu \nu}}{\partial x^{0}}=0$, where $x^0$ is a timelike coordinate) \cite{d1992introducing}. Being stationary does not restrict the metric to have multiple phrases. To extricate these types of phrases one has to impose another condition on the metric as being static, which means that the metric is time-reversal invariant so multiple phrases will be omitted. Then one can apply spherical symmetry that means space-time has three spacelike Killing vector fields 
	 $X^{\alpha}$,
	with the following relation
	  \begin{equation}
	 [X^1,X^2]=X^3\quad ,\quad [X^2,X^3]=X^1 \quad,\quad [X^3,X^1]=X^2.
	 \end{equation}
	 Finally the metric take the simplified form
	 	\begin{equation}
	 ds^{2}=-e^{A(r)} dt^{2}+e^{B(r)} dr^{2}+r^{2} d {\theta}^{2}+r^{2} sin^{2}{\theta} d{\varphi}^{2}.
	 \label{eq:2.22}
	 \end{equation}
	  $A(r)$ has been replaced to $e^{A(r)}$ because metric's elements are always positive and this choice will simplify calculations.
	  \subsection{Ricci tensor calculation}
According to the line element
\eqref{eq:2.22}
one can obtain non-zero components of the Ricci tensor as bellow
\begin{align}
\label{eq:2.500}
&R_{tt} = {e} ^ {A-B} (1/2A''-1/4A'B'+1/4 {A'} ^ {2} +A'/r),\\
&R_{rr} =-1/2A''+1/4A'B'-1/4 {A'} ^ {2} +B'/r,\notag\\
&R_{\theta \theta } =- {e} ^ {-B}  (1+ \frac {r({A} ^ {'} - {B} ^ {'}  )} {2 })  +1,\notag\\
&R_{\varphi \varphi}=sin^{2}{\theta}R_{\theta \theta}.\notag
\end{align}
 Putting these relations in
	\eqref{eq:2.3} 
	 it is possible to write
	 \begin{equation}
	 \Lambda {e} ^ {A} = {e} ^ {A-B} (1/2A''-1/4A'B'+1/4 {A'} ^ {2} +A'/r)+2\Lambda e^{A}
	 \label{eq:2.8}
	 \end{equation}
and
	 \begin{equation}
	 -\Lambda {e} ^ {B} =-1/2A''+1/4A'B'-1/4 {A'} ^ {2} +B'/r-2\Lambda e^{B}.
	 \label{eq:2.9}
	 \end{equation}
	 Dividing
	 \eqref{eq:2.8}
by
	 \eqref{eq:2.9} 
	 one has
	 \begin{equation}
	 A'=-B'\quad,\quad A=-B,
	 \end{equation}
	 where the integral constant is considered to be zero.
	 If one puts the third relation of \eqref{eq:2.500}
in
	 $R_{\mu \nu}=\Lambda g_{\mu \nu}$
	 then
	 \begin{align}
	 &e^{A}(1+rA')=1-\Lambda r^{2}\\
	 &X\equiv e^{A(r)}\notag\\
	 &X+rX'=1-\Lambda r^{2}\notag\\
	 &\frac{d}{dr}(rX)=\frac{d}{dr}(r-(\Lambda/3)r^{3})\notag\\
	 &rX=r-(\Lambda/3)r^{3}+C,\notag
	 \end{align}
	 where $C$ is the integral constant. Considering Newtonian limit, a matter field at the point $O$ causes the potential $\phi=-\frac{GM}{r}$ \cite{d1992introducing}. This potential in weak field limit results 
	 \begin{equation}
	 g_{00}\simeq 1+2\phi/c^2=1-2GM/c^2r,
	 \end{equation}
	 therefore
	 $C\equiv GM/c^2$. Considering
	 $c$
	 and
	 $G$
	 to be equal to one, $e^{A}$ can be obtained as follows
	 \begin{equation}
	 e^{A}=1-(\Lambda /3) r^{2}+2M/r.
	 \end{equation} 
	 Putting this relation in \eqref{eq:2.22} the line element becomes
	 \begin{equation}
	 {ds} ^ {2} =-  (1-\frac {2M} {r} - {\Lambda} \frac {{r} ^ {2}}{3}   ) {dt} ^ {2} + { (1- \frac{2M} {r} - {\Lambda}  \frac {{r} ^ {2}}{3}   )} ^ {-1} {dr} ^ {2} + {r} ^ {2} {d \Omega} ^ {2}.
	 \label{eq:2.14}
	 \end{equation}
	Actually, this is de Sitter-Schwarzschild solution. If $\Lambda=0$ in equation \eqref{eq:2.14} then the 
	 Schwarzschild solution for Einstein equations will be acquired and if $M=0$  de Sitter solution will be obtained
	 \begin{equation}
	 {ds} ^ {2} =-  (1 - {\Lambda} \frac {{r} ^ {2}}{3}   ) {dt} ^ {2} + { (1 - {\Lambda}  \frac {{r} ^ {2}}{3}   )} ^ {-1} {dr} ^ {2} + {r} ^ {2} {d \Omega} ^ {2}.
	 \label{eq:2.18}
	 \end{equation}
	 Let $l\equiv\sqrt{3/ \Lambda}$  and \eqref{eq:2.18} becomes
	 \begin{equation}
	 {ds} ^ {2} =-  (1 - \frac {{r} ^ {2}}{l^{2}}   ) {dt} ^ {2} + { (1 -   \frac {{r} ^ {2}}{l^{2}}   )} ^ {-1} {dr} ^ {2} + {r} ^ {2} {d \Omega} ^ {2}.
	 \label{eq:2.29}
	 \end{equation}
	 This is de Sitter line element.
	 \section{De Sitter horizon}
	 If  $r=l$  then components of the line element
	 \eqref{eq:2.29} become  irreversible (caused by the choice of coordinates) and  a singularity appears in this value of $r$.  The difference between this horizon and the Schwarzschild's horizon is that in de Sitter space-time, each observer has his/her own unique horizon.  As a result
	  $t,r,\theta,\varphi$
	  are not appropriate coordinates to describe the whole de Sitter manifold. More precisely they are unable to delineate beyond the horizon. Therefore this coordinate system do not describe the whole de Sitter space-time, it can only explain the static patch of de Sitter space-time \cite{hawking1973large}. Afterwards it is useful to find other coordinate systems that satisfy our aim to study beyond the horizon. 
	 \section{Embedding de Sitter space-time in five dimensional Minkowski space-time\label{sec:2.2.2} }
	 Different coordinates are describable for each space-time. De Sitter space-time is no exception. Therefore one can describe various coordinates for it. One can defines $(x_0,x_1,x_2,x_3,x_4)$ as follows
	 \begin{align}
	 \label{eq: embedd}
	 &t=l\tanh^{-1}(x_0/x_1),\\
	 &r=\sqrt{x_0^{2}-x_1^{2}+l^2}\notag,\\
	 &\theta=\cos^{-1}(\frac{x_4}{\sqrt{x_0^{2}-x_1^{2}+l^2}}),\notag\\
	 &\varphi=\tan(x_3/x_4)\notag.
	 \end{align}
	 If one puts this relations in 
	 \eqref{eq:2.29}
	 the following relation will be obtained
	 \begin{equation}
	 ds^{2}=-dx^{2}_{0}+dx^{2}_{1}+dx^{2}_{2}+dx^{2}_{3}+dx^{2}_{4},
	 \label{eq:(2.25)}
	 \end{equation}
	\eqref{eq: embedd} can be written as \cite{pascu2012atlas}
	 \begin{align}
	 	\label{eq:2.30}
	&{x}_0 =l \sqrt{ (1-\frac{{r}^{2}}{{l}^{2}}  )} \sinh(\frac{t}{l} ),\\
	&{x}_{1} =l \sqrt{ (1-\frac{{r}^{2}}{{l}^{2}})} \cosh(\frac{t}{l}),\notag\\
	&x_{2}=r \sin{\theta}\cos{\varphi},\notag\\
	&x_{3}=r \sin{\theta}\sin{\varphi},\notag\\
	&x_{4}=r \cos{\theta}.\notag
	\end{align}
by squaring these components and sum them one can writes 
\begin{equation}
-x^{2}_{0}+x^{2}_{1}+x^{2}_{2}+x^{2}_{3}+x^{2}_{4}=l^{2}.
\label{eq:(2.26)}
\end{equation}
This is the relation of a hyperboloid embedded in five dimensional Minkowski space-time. This relation is analogues to  \eqref{eq:2.200}. Hence relation \eqref{eq:2.200} can also be obtained by solving Einstein equations.
\section{De Sitter hyperbloid}
Another appropriate coordinates to designate de Sitter space-time is
\begin{align}
\label{eq:2.32}
&x_{0}=l\sinh (\tau/l),\\
&x_{1}=l\cosh (\tau/l)\cos(\theta),\notag\\
&x_{2}=l\cosh (\tau/l)\sin(\theta)\cos(\varphi),\notag\\
&x_{3}=l\cosh (\tau/l)\sin(\theta)\sin(\varphi)\cos(\alpha),\notag\\
&x_{4}=l\cosh (\tau/l)\sin(\theta)\sin(\varphi)\sin(\alpha),\notag
\end{align} 
	 where $\tau=l \sinh^{-1}[l \sqrt{1-(r/l)^{2}}\sinh(t/l)]$.
	 
\begin{figure}
	   	\centering
	   	\includegraphics[width=50mm]{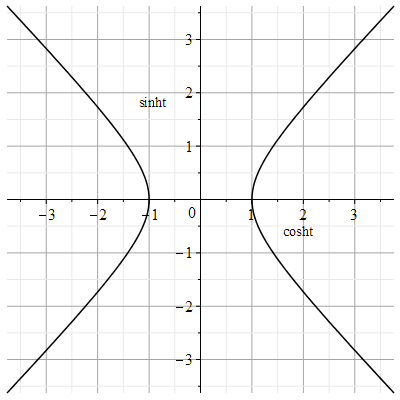}
	   	\caption{ This figure illustrates 
	   		$\cosh^{2}(\tau/l)-\sinh^{2}(\tau/l)=1$.
	   	  }
	   	\label{fig:2.33}
	   \end{figure}

	  If one puts 
	   \eqref{eq:2.32}
	   components in \eqref{eq:(2.26)} 
	   then a common equation will appear
	   \begin{equation}
	   \cosh^{2}(\tau/l)-\sinh^{2}(\tau/l)=1,
	   \label{eq:2.44}
	   \end{equation}
	   with the form \ref{fig:2.33}.
	 
\begin{figure}
	 	\centering
	 	\includegraphics[width=70mm]{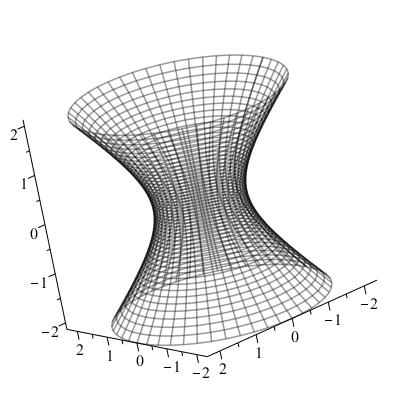}
	 	\caption{De Sitter hyperbloid that shows global coordinates. }
	 	\label{fig:(2.1)}	
	 \end{figure}

	 On the other hand for spacelike part we have the relation of a two-sphere. Then for the whole space-time one can see \ref{fig:(2.1)}
 \begin{figure}
 	\centering
 		\includegraphics[width=70mm]{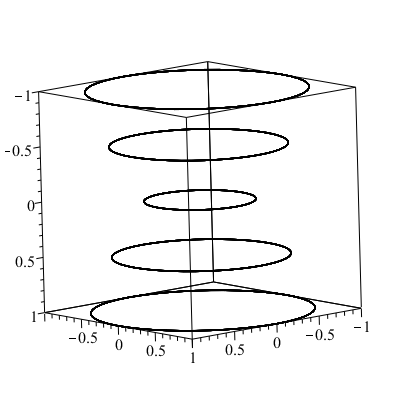}
 		\caption{In this figure circles show surfaces of constant t.}
 \end{figure}
\begin{figure}
	 	\centering
 	\includegraphics[width=70mm]{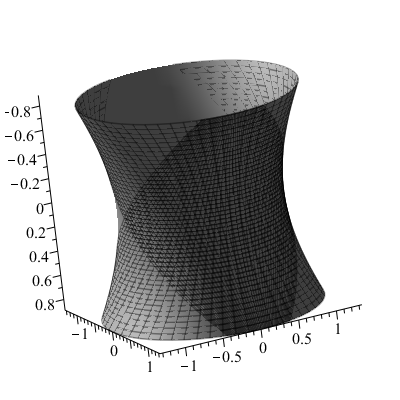}
 	\caption{Shaded region shows static de Sitter coordinates}
 \end{figure}
\begin{figure}
	 	\centering
	\includegraphics[width=70mm]{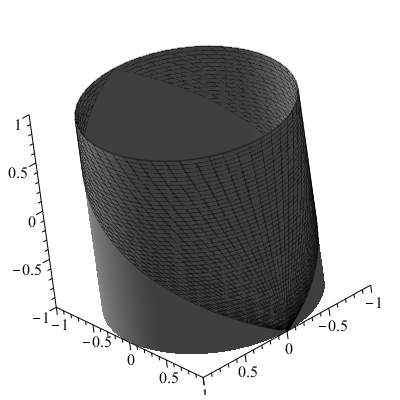}
	\caption{De Sitter space-time is conformal to the part $-\pi/2<T<\pi/2$ of the Einstein static universe.}
 \end{figure}
\section{Global coordinates}
As it was explained in section \ref{sec:2.2.2} one can defines various coordinates with respect to the equation \eqref{eq:(2.26)}. Therefore it is possible to write
\begin{align}
\label{eq:2.36} 
&x_{0}=l \sinh(\tau /l),\\
\label{eq:2.37}  
&x_{j}=l \cosh(\tau /l) \omega^{j}.  
\end{align}
Where $(j=1,2,3,4)$. Then the line element takes the form
 \begin{equation}
ds^{2}=-d \tau^{2}+l^{2}\cosh^{2}(\tau/l) d \omega^{(j)2}.
\label{eq:2.33}
\end{equation} 

	A conformal factor can be multiply in \eqref{eq:2.33} and the compactified space-time (see Appendix \ref{app:A} ) can be obtained.

	 \section{De Sitter space-time's completion}
	 Talking about infinity is not an easy task, because it is vast and out of reach \cite{23}. Hence with aid of compactification infinity becomes accessible (see Appendix \ref{app:A}). One can multiply the conformal factor
	 	 $\Omega=\frac{1}{l^{2}\cosh^{2}(\tau/l)}$ 
	 	in the metric 
	 	\eqref{eq:2.33}. Causalty is invariant under conformal transformation although it changes geometry.   Geometry's change allows us to have infinities in the problem as well. Multiplying this conformal factor in 
	 	 \eqref{eq:(2.25)} 
	 	 one can see that de Sitter space-time is locally conformal to the Einstein static universe
	 \begin{equation}
	ds^{2}=l^{2}\cosh^{2}(\tau/l)d\bar{s}^{2}.
	\end{equation}
	Where
	$d\bar{s}^{2}$
	is the line element of Einstein static universe.
\begin{align}
&d\bar{s}^{2}=-l^{-2}\cosh^{-2}(\tau/l)d\tau^{2}+dR^{2}+R^{2}d \Omega^{2},\\
& \xrightarrow{T=\tau/l\cosh^{-1}(\tau/l)-\sqrt{\tau^2/l^2-1}}\notag\\
&d \bar{s}^{2}=-dT^{2}+dR^{2}+R^{2}d \Omega^{2}.\notag
\end{align}
	 Considering the range of this components one can find infinities for this nonphysical metric.
	 De Sitter space-time unlike Minkowski space-time  has spacelike infinity for null and timelike lines.

	 \section{Killing vector fields}
	 A vector field has conformal motions if \cite{hassani2001mathematical} 
	 \begin{equation}
	 \mathcal{L}_{X}g_{\mu \nu}=2\phi(x^{\sigma})g_{\mu \nu}.
	 \end{equation}
	 In this equation if $\phi$ is constant then $X$ will be a homothetic vector and if $\phi=0$,  $X$ will be a Killing vector \cite{stephani2009exact}. 
	 
	 Thus according to $\mathcal{L}_{X}g_{\mu \nu}=0$  Killing vector fields for de Sitter space-time are written as follows \cite{banerjee2007gauge,salcedo2017sitter,yan2017killing}
	 
	 \begin{align}
	 &\xi^{\mu}=  [\frac{r\cos(\theta)l\exp(t/l)}{\sqrt{(l^2-r^2)}},\cos(\theta)\exp(t/l)\sqrt{(l^2-r^2)}, \frac{-\exp(t/l)\sin(\theta)\sqrt{(l^2-r^2)}}{r}, 0] ,\\
	 &\xi^{\mu}=    [0, 0, 0,1]   ,\notag\\
	 &\xi^{\mu}=    [\frac{r\sin(\theta)\sin(\phi)l\exp(t/l)}{\sqrt{(l^2-r^2)}},\sin(\theta)\sin(\varphi)\exp(t/l)\sqrt{(l^2-r^2)},\notag\\   &\frac{\exp(t/l)\cos(\theta)\sin(\varphi)\sqrt{(l^2-r^2)}}{r},\frac{\exp(t/l)\cos(\varphi)(l-r)(l+r)}{\sqrt{(l^2-r^2)r\sin(\theta)}}]   ,\notag\\
	 &\xi^{\mu}= \frac{r\sin(\theta)\cos(\varphi)l\exp(t/l)}{\sqrt{(l^2-r^2)}},  [\sin(\theta)\cos(\varphi)\exp(t/l)\sqrt{(l^2-r^2)},\notag\\   &\frac{\exp(t/l)\cos(\theta)\cos(\varphi)\sqrt{(l^2-r^2)}}{r},\frac{-\exp(t/l)\sin(\varphi)(l-r)(l+r)}{\sqrt{(l^2-r^2)}r\sin(\theta)}]    ,\notag\\
	 &\xi^{\mu}=   [\frac{-rl\cos(\theta)\exp(-t/l)}{\sqrt{(l^2-r^2)}},\cos(\theta)\exp(-t/l)\sqrt{(l^2-r^2)}, \frac{-\sin(\theta)\exp(-t/l)\sqrt{(l^2-r^2)}}{r}, 0]    ,\notag\\
	 &\xi^{\mu}= [-rl\sin(\theta)\sin(\varphi)\exp(-t/l)/\sqrt{(l^2-r^2)},\sin(\theta)\sin(\varphi)\exp(-t/l)\sqrt{(l^2-r^2)},\notag\\  &\frac{\cos(\theta)\sin(\varphi)\exp(-t/l)\sqrt{(l^2-r^2)}}{r},\frac{\cos(\varphi)\exp(-t/l)(l-r)(l+r)}{\sqrt{(l^2-r^2)}r\sin(\theta)}] ,\notag\\
	 &\xi^{\mu}=[\frac{-rl\sin(\theta)\cos(\varphi)exp(-t/l)}{\sqrt{(l^2-r^2)}},\sin(\theta)\cos(\varphi)\exp(-t/l)\sqrt{(l^2-r^2)},\notag\\
	 &\frac{\cos(\theta)\cos(\varphi)exp(-t/l)\sqrt{(l^2-r^2)}}{r}, \frac{-\sin(\varphi)\exp(-t/l)(l-r)(l+r)}{\sqrt{(l^2-r^2)}r\sin(\theta)}]       ,\notag\\
	 &\xi^{\mu}=   [1, 0, 0, 0]    ,\notag\\
	 &\xi^{\mu} = [0, 0,\sin(\varphi), \frac{\cos(\varphi)}{\tan(\theta)}]     ,\notag\\
	 &\xi^{\mu}=    [0,0, \cos(\varphi), -\sin(\varphi)/\tan(\theta)]   ,\notag
	 \end{align}
	 As one expected for de Sitter space-time that is maximally symmetric, ten killing vector fields have been found.
	 \chapter{De Sitter space-time's symmetries}
	 Symmetry in physics is a mathematical or physical property of a system that remains unchanged under certain transformations. There exists quantities which are expected to be   invariant under transformations, so we look for symmetry groups that hold these quantities invariant. In the following, we will talk about the existence of such a quantity for gravitational systems and the symmetry group that maintains its inefficiency.
	 \section{Planck length \label{sec:4.1}}
	  Planck length, $l_p$, is the distance that light travels in Planck time. This length can be described by three fundamental physical constants, speed of light in vacuum, $c$, Planck constant, $h$, gravitational constant, $G$. Though its relation is
	  \begin{equation}
	  l_p=\sqrt{\frac{\hbar G}{c^3}}=1/616229(38)\times 10^{-35} m.
	  \label{eq:3.1}
	  \end{equation}
	  In 1899 Max Planck proposed to use certain constants for length, mass, time and energy. Considering only the Newton gravitational  constant, speed of light and Planck constant he found these constants named Planck length, Planck mass, Planck time and Plank energy. 
	  
	  Quantum effects are believed to appear in this scale. to measure anything in this scale the photon momentum must be very high. Considering Heisenberg's uncertainty principle a black hole appears in this scale that its horizon is equal to Planck length. One can rewrite the uncertainty principle as
	  \begin{equation}
	  \Delta p\Delta r>\hbar/2.
	  \end{equation}
	  multiplying both sides by 	$2\frac{G }{c^3}$ one gets \cite{aurilia2013planck,carr2016black}
	  \begin{align}
	  \label{eq:3.444}
	  &\Delta(\frac{2Gm}{c^2})\Delta r> \frac{G \hbar}{c^3}\\
	  \Rightarrow&\Delta r_s \Delta r>l_p^2.\notag
	  \end{align}
	  where $r_s$  is gravitational radius, $r$  is coordinates radius and $l_p$  is Planck length. Relation  \eqref{eq:3.444} is the uncertainty principle in  quantum gravity. 
	  
	  Uncertainty principle anticipate the existence of black holes and wormholes so any attempt to acquire a distance smaller than Planck length is considered impossible because a black hole will appear in this distance \cite{carr2016black}.
	  
	  The Lorentz symmetry group does not remain invariant at this minimum length. Note that in special relativity on the flat background, the closer the speed to the speed of light, the closer the length goes to zero
	  	\begin{equation}
	  L=L_0\sqrt{1-v^2 / c^2}.
	  \end{equation}
	  This is in contrast to the invariance of the Planck length.
	  
	 Lorentz group can be only  realized on homogeneous space-time that means except Minkowski space-time it can be written on de Sitter and anti-de Sitter space-times that are the only possible homogeneous  space-times  in $(3+1)$ dimensions. In this thesis we focus on de Sitter space-time with constant positive scalar curvature.
	
	 	\begin{equation}
	 R=12 l^{-2}.
	 \label{eq:3.4}
	 \end{equation}
  Where $l$ is the de Sitter length.  Equation  \eqref{eq:3.4} shows the relation between Ricci scalar and length. We know by the definition that Lorentz transformations will remain the curvature invariant. Thus Lorents transformations on de Sitter space-time remain de Sitter length invariant \cite{aldrovandi2007sitter,araujo2017sitter,araujo2019sitter,gaarder2016gravitational,gibbons2003newton, salcedo2017sitter}. Somehow, we also have this concept hidden in Minkowski space-time, what remains invariant there is an infinite  length and does not affect the space-time's curvature.
	 \section{De Sitter transformations}
	 De Sitter transformations can be known as rotations in five dimensional Euclidean space. Each observer has his/her own coordinate set, we want to find transformations between these coordinate sets. We search for a symmetry group that remains the metric invariant. This group named as de Sitter group $So(4,1)$ that apply in the embedding five dimensional Minkowski space-time
	 \begin{equation}
	 X^{\sigma}=\Lambda^{\sigma}_{\rho}X^{\rho}
	 \end{equation}
	 where $\Lambda^{\sigma}_{\rho}$ is the group element. In vector representation one has \cite{hartman2017lecture}
	 \begin{equation}
	 -g_{\mu \nu}X^{\mu}X^{\nu}=l^2
	 \end{equation}
	 that shows these transformations remain the length invariant. Infinitesimal transformations can be shown in the following relation
	 \begin{equation}
	 \delta X^{\sigma}=1/2 \xi^{\mu \nu} L_{\mu \nu} X^{\sigma}
	 \label{eq:3.12}
	 \end{equation}
	 where
	 $L_{\mu \nu}$
	 and
	  $\xi^{\mu \nu}$ 
	 are generators and parameters of de Sitter translations.
	 \section{Spherical rotations\label{3.3}}
	 According to the equation 	\eqref{eq:2.30}
	 one can see that the last three components are parameterize in 	$ \mathbb{R}^3$ then symmetry elements  are rotation elements
	 \begin{equation}
	 \mathcal{G}_{rot}^{(1)}=
	 \begin{bmatrix}
	 1&0&0&0&0\\
	 0&1&0&0&0\\
	 0&0&1&0&0\\
	 0&0&0&\cos{\alpha}&-\sin{\alpha}\\
	 0&0&0&\sin{\alpha}&\cos{\alpha}
	 \end{bmatrix},
	 \end{equation}
	 \begin{equation}
	 \mathcal{G}_{rot}^{(2)}=
	 \begin{bmatrix}
	 1&0&0&0&0\\
	 0&1&0&0&0\\
	 0&0&\cos{\alpha}&-\sin{\alpha}&0\\
	 0&0&\sin{\alpha}&\cos{\alpha}&0\\
	 0&0&0&0&1
	 
	 \end{bmatrix},
	 \end{equation}
	 \begin{equation}
	 \mathcal{G}_{rot}^{(3)}=
	 \begin{bmatrix}
	 1&0&0&0&0\\
	 0&1&0&0&0\\
	 0&0&\cos{\alpha}&0&-\sin{\alpha}\\
	 0&0&0&1&0\\
	 0&0&\sin{\alpha}&0&\cos{\alpha}
	 \end{bmatrix}.
	 \end{equation}
	 \section{Time translation}
	 As de Sitter metric is static then it should be invariant under time translation
	 	\begin{equation}
	 \mathcal{T}_{trans}^{(1)}=
	 \begin{bmatrix}
	 \cosh{(\beta /l)}&-\sinh{(\beta /l)}&0&0&0\\
	 -\sinh{(\beta /l)}&\cosh{(\beta /l)}&0&0&0\\
	 0&0&1&0&0\\
	 0&0&0&1&0\\
	 0&0&0&0&1
	 
	 \end{bmatrix}.
	 \end{equation}
	 This transformation can be considered as a boost in $x_1$ direction.
	 \section{Rotations on the hyperboloid}
	 As one can see in section \ref{3.3} for spherical rotation the $x_1$ axis is considered a constant axis. Afterwards allowing $x_1$ axis to be variable another subgroup of rotation known as  rotations on the hyperboloid appear
	 \begin{equation}
	 \mathcal{R}_{rot}^{(1)}=
	 \begin{bmatrix}
	 1&0&0&0&0\\
	 0&\cos{\alpha}&-\sin{\alpha}&0&0\\
	 0&\sin{\alpha}&\cos{\alpha}&0&0\\
	 0&0&0&1&0\\
	 0&0&0&0&1
	 \end{bmatrix},
	 \end{equation}
	 \begin{equation}
	 \mathcal{R}_{rot}^{(2)}=
	 \begin{bmatrix}
	 1&0&0&0&0\\
	 0&\cos{\alpha}&0&-\sin{\alpha}&0\\
	 0&0&1&0&0\\
	 0&\sin{\alpha}&0&\cos{\alpha}&0\\
	 0&0&0&0&1
	 \end{bmatrix},
	 \end{equation}
	 \begin{equation}
	 \mathcal{R}_{rot}^{(3)}=
	 \begin{bmatrix}
	 1&0&0&0&0\\
	 0&\cos{\alpha}&0&0&-\sin{\alpha}\\
	 0&0&1&0&0\\
	 0&0&0&1&0\\
	 0&\sin{\alpha}&0&0&\cos{\alpha}
	 \end{bmatrix}.
	 \end{equation}
	 \section{boosts}
	 Other transformations that we have to consider are boosts
	 \begin{equation}
	 \begin{bmatrix}
	 \cosh{\beta}&0&-\sinh{\beta}&0&0\\
	 0&1&0&0&0\\
	 -\sinh{\beta}&0&\cosh{\beta}&0&0\\
	 0&0&0&1&0\\
	 0&0&0&0&1
	 \end{bmatrix},
	 \end{equation}
	 \begin{equation}
	 \begin{bmatrix}
	 \cosh{\beta}&0&0&-\sinh{\beta}&0\\
	 0&1&0&0&0\\
	 0&0&1&0&0\\
	 -\sinh{\beta}&0&0&\cosh{\beta}&0\\
	 0&0&0&0&1
	 \end{bmatrix},
	 \end{equation}
	 \begin{equation}
	 \begin{bmatrix}
	 \cosh{\beta}&0&0&0&-\sinh{\beta}\\
	 0&1&0&0&0\\
	 0&0&1&0&0\\
	 0&0&0&1&0\\
	 -\sinh{\beta}&0&0&0&\cosh{\beta}
	 \end{bmatrix}.
	 \end{equation}
	 \section{Conformal transformations}
	 Number of group elements for the group $SO(1,n+1)$ can be obtained according to the following relation
	 \begin{equation}
	 dimSO(1,n+1)=\frac{1}{2}(n+1)(n+2).
	 \label{eq:3.5}
	 \end{equation}
	 As an example Pioncaré group in $n$ dimensions has $n$ translation generators and $\frac{n(n-1)}{2}$ rotation generators.
	 \begin{equation}
	 dim Poincar \acute{e}(E^{n})=\frac{1}{2}n(n+1).
	 \label{eq:3.6}
	 \end{equation}
	 These calculations have been done locally. De Sitter space-time is maximally symmetric so having the curvature of a point, one can find the  space-time's curvature so results of
	 \eqref{eq:3.5}
	 and
	 \eqref{eq:3.6}
	  can be referred to the whole space-time.
	  The deference between 
	 \eqref{eq:3.5}
	 and 
	 \eqref{eq:3.6} is $n+1$. This incompatibility can be described by adding conformal transformations. In the following different types of these transformations are represented \cite{duval2011conformal}.
	 
	 Multiplying a conformal factor to a vector one obtains a transformations named as dilations 
	 	\begin{equation}
	 \vec{x}\rightarrow \lambda \vec{x}\quad,\quad\lambda \in \mathbb{R}.
	 \end{equation}
	 with
	 \begin{equation}
	 D=t\partial_t+x\partial_x+y\partial_y+z\partial_z
	 \end{equation}
	 as their generator.
	 
	 Other relevant type of transformations are conformal spatial transformations
	 $\vec{x} \rightarrow \vec{x}'$, as
	 \begin{equation}
	 \frac{x'^{\mu}}{x'^{2}}=\frac{x^{\mu}}{x^{2}}+\alpha^{\mu},
	 \end{equation}
	 where $x^{2}=x_{\mu}x^{\mu}$ and $\mu=1,\dots,n$.
	 One can also write
	 \begin{equation}
	 x'^{\mu}=\frac{x^{\mu}+\alpha^{\mu}x^{2}}{1+2\alpha_{\mu}x^{\mu}+\alpha^{2}x^{2}},
	 \end{equation}
	 with four generators
	 \begin{align}
	 &K_1=(t^2+x^2+y^2+z^2)\partial_t+2xt\partial_x+2yt\partial_y+2zt\partial_z,\\
	 &K_2=2xt\partial_t+(t^2+x^2+y^2+z^2)\partial_x+2ty\partial_y+2zt\partial_z,\notag\\
	 &K_3=2ty\partial_t+2xt\partial_x+(t^2+x^2+y^2+z^2)\partial_y+2zt\partial_z,\notag\\
	 &K_4=2tz\partial_{t}+2xt\partial_x+2yt\partial_y+(t^2+x^2+y^2+z^2)\partial_z.\notag
	 \end{align}
	 As one can see in previous pages, we have ten generators for rotations, time translation and boosts and five generators for conformal transformations so we have $SO(1,4)$ as de Sitter symmetry group.
	 \section{Commutation relations}
	 It is possible to study the Lie algebra for conformal transformations (see table \ref{tab 3.1}).
	 \begin{table} 	
	 	\begin{tabular}{|p{2.5cm}|p{5cm}|p{2.5cm}|p{5cm}|}
	 		\hline
	 	Translations&Rotations&	Dilations&Conformal spatial transformations \\ \hline
	 	$P_{\mu}=-\partial_{\mu}$&$M_{\mu\nu}=(x_{\mu}\partial_{\nu}-x_{\nu}\partial_{\mu})=(P_{\mu}\partial_{\nu}-P_{\nu}\partial_{\mu})$&$D=-x^{\mu}\partial_{\mu}$&$K_{\mu}=(2x_{\mu}x^{\nu}\partial_{\nu}-x^{2}\partial_{\mu})=-2x_{\mu}D+x^{2}P_{\mu}$\\ \hline
	 	\end{tabular}
 	\caption{So(4,1) generators. \label{tab 3.1}}
 	 \end{table}
 	So one can write the algebra that rules on these transformations
 	\begin{align}
 	\label{eq:3.54}
 	&[M_{\mu\nu},P_{\rho}]=(g_{\nu\rho}P_{\mu}-g_{\mu\rho}P_{\nu}),\\
 	&[M_{\mu\nu},M_{\rho \tau}]=(g_{\mu \tau}M_{\nu \rho}+g_{\nu \rho}M_{\mu \tau}-g_{\mu\rho}M_{\nu\tau}-g_{\nu\tau}M_{\mu\rho}),\notag\\
 	&[M_{\mu \nu},K_{\rho}]=(g_{\nu \rho}K_{\mu}-g_{\mu\rho}K_{\nu}),\notag\\
 	&[D,P_{\mu}]=+P_{\mu},\notag\\
 	&[D,K_{\mu}]=+K_{\mu},\notag\\
 	&[P_{\mu},K_{\nu}]=2(g_{\mu \nu}D+M_{\mu \nu}).\notag
 	\end{align}
 	Other combinations are zero.
 	\chapter{Asymptotic symmetries}
 	De Sitter space-time's symmetries have been considered in the previous chapter. In the following chapter we are interested in  obtaining asymptotic symmetries. Our focus is on null infinity, $\mathcal{I}$, thus we are facing two problems ahead. First, according to the compactification, the topology changes, then the topology of  $\mathcal{I}$ is quite different from the topology of the physical space-time. Therefore, we can not necessarily say that the ruling symmetry group at null infinity is the same as the one at physical space-time. This subject has been studied in Minkowski space-time, more details on this method will follow. But there is another important matter that
 	  $\mathcal{I}$
 	  is null in asymptotically flat space-times and spacelike in asymptotically de Sitter space-times. Therefore  $\Lambda\rightarrow0$ does not have a continuous limit. This fact has important consequences.
 	  
 	  In this chapter these two issues are considered  and  useful methods to find asymptotic symmetries for de Sitter space-time are presented.
 	  \section{General discussion} 
 	  Talking about infinity is not facile so with the help of compactification (see Appendix \ref{app:A}) infinity becomes more palpable. As we said before by multiplying a conformal factor by the metric ($\tilde{ g}_{\mu\nu}=\Omega^2{ g}_{\mu\nu}$), one manage to attribute infinity to the boundary of a larger space-time, $(\tilde{M},\tilde{ g}_{\mu\nu})$. Conformal factor does not change the causality but it changes the geometry therefore we can not distinguish the symmetry group that rules the infinity without precise consideration. Actually the only phrase that one can say is that the group of diffeomorphisms is the appropriate symmetry group  which is not useful, because one can not define preserve charge according to them. In fact, diffeomorphism invariance is a local symmetry while we need a global symmetry to describe Noether charge.
 	  
    Using our knowledge about  physical space-time's properties and considering their transition when $r\rightarrow \infty$ might be a good way to find features of the infinity.
As said before, conformal transformation does not change the causality. Hence it is possible to consider gravitational fields and their asymptotic limits. As gravitational fields move with the speed of light, studying null infinity and finding a useful notion for it  may be feasible according to them.
 	  
 	  At first we will review the asymptotic behavior of a gravitational field in an isolated system. Actually this case is much easier than the other ones. Observations show that a system with a positive cosmological constant is more appropriate to describe our universe. Unfortunately one can not  use the process that is used in $\Lambda =0$ cases, in 
 	    $\Lambda >0$  cases \cite{bros2002asymptotic}.
 	    
 	    It is difficult to study the asymptotic structure of a gravitational field because the field itself changes the geometry of space-time. This issue becomes clear after the work of 
 	    Arnowitt, Deser and Misner at spacelike infinity \cite{arnowitt2008republication} and the work of Bondi, Sachs and Newman at null infinity \cite{bondi1962gravitational}. In ADM framework space-time is divided to time constant surfaces. Each surface has a three dimensional metric $\lambda_{ij}(t,x^k)$ and a momentum $\pi^{ij}(t,x^k)$ according to that one can define the Hamiltonian (see \cite{arnowitt2008republication,deser1967covariant} ).
 	    
 	    Now let's talk about null infinity, our main subject. First we will review the work of Bondi and his collaborators. They established a system for studying the expansion of the metric on a null path. Null infinity,  $\mathcal{I}$, is known as the boundary for physical space-time. considering this gravitational radiation on $\mathcal{I}$ one can defines the Bondi news,
 	    $N_{\mu\nu}$,
 	    which in
 	    Bondi–Sachs physical space coordinates, $\hat{x}^{\mu}=(u,l,x^{\mu})=(t-r,1/r,x^{\mu})$,
 	    has the form  \cite{ashtekar2014asymptotics}
 	    \begin{equation}
 	    N_{\mu\nu}=\zeta^*(\lim_{l\rightarrow0}l^{-1}\hat{\nabla}_{\mu}\hat{\nabla}_{\nu}l),
 	    \end{equation}
 	    where $\zeta^*$ shows the pullback to $\mathcal{I}^+$. Two components of $N_{\mu\nu}$ show the two possible modes in exact general relativity. In asymptotically flat space-times if and only if 
 	     $N_{\mu\nu} \neq 0$,
 	     gravitational radiation do not carry energy momentum on  $\mathcal{I}$. If $N_{\mu \nu} \neq 0$, 
 	     $\eta_{\mu \nu}$  is no longer unique and one has to defines a new metric, $\eta'_{\mu \nu}$ with the translation $t \rightarrow t'=t+f(\theta,\varphi)$ as $g_{\mu \nu}$ has the same asymptotic behavior with it. Thus the asymptotic symmetry group is not Pioncaré group but another group, called as BMS group including an abelian subgroup $\mathfrak{T}$ that contains translations just like the subgroup that has been defined for Pioncaré group. Hence the definition of energy momentum at null infinity is well-defined \cite{bondi1962gravitational}.
 	     
 	     For the $\Lambda>0$ case the process is quite different. As  $\mathcal{I}$ is a spacelike hypersurface one can not obtain the symmetry group the way the $BMS$ group has been obtained. It should be added that by considering $1/r$ coordinate transformation for a locally de Sitter Bondi-Sachs metric can be used and a symmetry group named as $\Lambda-BMS$ group has been obtained
 	     To study such a structure, we first need to consider the basic definitions \cite{poole2019ds4,compere2019advanced,aneesh2019conserved}.
 	     \section{Asymptotically flat space-time}
 	     In physics we like to study isolated systems. If a space-time becomes flat when $r\rightarrow \infty$, this space-time  is asymptotically flat and asymptotically flat space-times are isolated \cite{wald2000general,calo2018relation}.
 	     
 	     Finding a meaningful definition for isolated systems in general relativity is not simple because finding a helpful description for infinity is difficult \cite{wald1999gravitational}. Compactifying the space-time (see Appendix \ref{app:A}) is a useful method to have a good definition for infinity. According to that, a definition for asymptotically flat space-time has also been found. A space-time is asymptotically flat if its null and space-like infinities become like the null and space-like infinities of the flat space-time. More precisely the space-time  $(M,g_{\mu\nu})$ is  asymptotically flat if a conformal space-time,  $(\tilde{M},\tilde{g}_{\mu\nu})$ exist as  $\tilde{g}_{\mu\nu}$ becomes $C^{\infty}$  everywhere except $i^0$ where it becomes  $C^{>0}$ and conformal isometrie, $\psi:M \rightarrow \psi[M]\subset \tilde{M}$ with conformal factor $\Omega$ as ${g}_{ab}=\Omega^{2} \psi^*\tilde{ g}_{ab}$ satisfy the following conditions.\cite{wald2000general}
 	     
 	     \noindent\fbox{
 	     	\parbox{\textwidth}{
 	     		1. $\bar{J^{+}}(i^0)\cup\bar{J^-}(i^0)=\tilde{M}-M$.
 	     		
 	     		2. There exists an open neighborhood, V, of $\mathring{M}= i^{0} \cup \mathcal{I}^{-}\cup \mathcal{I}^{+}$, where $(V,\tilde{g}_{\mu\nu})$ is strongly causal. 
 	     		
 	     		3. $\Omega$ can be extended to a function on all of the $\tilde{M}$ which is $C^{2}$ at $i^{0}$ and $C^{\infty}$ elsewhere.
 	     		
 	     		4. (a) For $\Omega$ at $\mathcal{I}$ one has
 	     		 \begin{align}
 	     		\label{eq:4.22222}
 	     		&\Omega|_{\mathcal{I}}=0,\\
 	     		&\tilde{\nabla}_{\mu} \Omega|_{\mathcal{I}}\neq 0,\notag
 	     		\end{align} 
 	     		where $\tilde{\nabla}_{\mu}$  is the covariant derivative according to $\tilde{g}_{\mu \nu}$.
 	     	
 	     		(b) At $i^0$ one can write
 	     		\begin{align}
 	     		&\Omega|_{i^{0}}=0,\\
 	     		&\lim_{i^{0}} \tilde{\nabla}_{\mu}\Omega=0,\notag\\
 	     		&\lim_{i^{0}} \tilde{\nabla}_{\mu}\tilde{\nabla}_{\nu}\Omega= 2 \tilde{g}_{\mu \nu}(i^{0}).\notag
 	     		\end{align}
 	     	}}
      	
      The condition \eqref{eq:4.22222} lets $\Omega$ to be a component on $\mathcal{I}$. One has the liberty to choose $\Omega$. Thus if one chooses the conformal frame $\tilde{\nabla}_{\mu}\tilde{n}^{\mu}|_{\mathcal{I}}=0$ it is possible to use $\tilde{n}^{\mu}$ as a component on the tangent space of the $\mathcal{I}$. This opportunity can be used to change the conformal scale as 
      	$\Omega \rightarrow \Omega'=\omega\Omega$, so
      	\begin{align}
      	&\tilde{n}'^{\mu}=\omega^{-1}\tilde{n}^{\mu},\\
      	&q'_{\mu\nu}|_{\mathcal{I}}=\omega^{2}q_{\mu\nu},\notag
      	\end{align} 
      	where $\mathcal{L}_{\tilde{n}}\omega=0$. Choosing  the conformal frame $\tilde{\nabla}_{\mu}\tilde{n}^{\mu}|_{\mathcal{I}}=0$, degrees of freedom decrease. Space-time in this conformal frame $q_{\mu\nu}$ has the signature $(0,+,+)$ at $\mathcal{I}$. 
      	\subsection{The Bondi Sachs metric}
       If one foliate the space-time to $u=constant$ hypersurfaces
The Bondi-Sachs coordinates $(u,r,x^A)$ can be used where $u=constant$ hypersurfaces  are null this imply that $g_{11}=0$ and we must have $\Gamma^{0}_{11}=\Gamma^{2}_{11}=0$ that results $r^4\sin^2\theta=g_{22}g_{33}$. The line element takes the form
      	\begin{align}
      	&ds^2=e^{2\beta}Vr^{-1}du^2-2e^{2\beta}du dr+r^2h_{AB}(dx^A-U^Adu)(dx^B-U^Bdu)
      	\end{align}
      	where $A,B=3,4$ and $\beta,\: V,\: h_{AB}$ are functions of $(u,\theta,\varphi)$. One can find asymptotic symmetries by checking all transformations that preserve this form of the line element \cite{bondi1962gravitational}, see also \cite{madler2016bondi}.
      	\section{Asymptotically flat space-times' symmetries}
       It is important to find symmetries that are presented with the vector $\xi^{\mu}
$ at null infinity. In other words near the equivalence class of vector fields that do not vanish at $\mathcal{I}$ \cite{chandrasekaran2018symmetries}. So it is possible to find vectors which satisfy the Killing equation near infinity. In curved space-time  there exists a large transformation group that depends on the angle and satisfies the Killing's equation. The asymptotic symmetry group,  $\mathfrak{G}$, defines as a quotient group as \cite{anninos2011asymptotic}
      	\begin{equation}
      	 \mathfrak{G}=Diff_{\infty}(\partial M)\setminus Diff^{0}_{\infty}(M),
      	\end{equation}
      	where  $Diff_{\infty}(M)$ is diffeomorphisms in physical space-time,  $({M},{g}_{\mu\nu})$ and  $Diff^{0}_{\infty}(M)$ is diffeomorphisms that are asymptotically identity. As said before, it is possible to use   $\tilde{n}^{\mu}$  as a component on  $\mathcal{I}$. When  $\mathcal{I}$ is null, $\tilde{n}^{\mu}$ lies on the tangent space of  $\mathcal{I}$ and with its aid one can define  $q_{\mu\nu}$ that has the signature  $(0,+,+)$. Field equations imply that  $\tilde{\nabla}_{\mu}\tilde{n}^{\mu}$ vanishes in each of these divergence-free conformal frames so the answers of the equation $\tilde{\nabla}_{\mu}\tilde{n}^{\mu}|_{\mathcal{I}}=0$ are the generators of the $\mathcal{I}$.
      	
       Actually  $BMS$ group is the symmetry group of $\mathcal{I}$. This group contains diffeomorphisms that remain the intrinsic metric, $q_{\mu\nu}$, and the vector field,  $n^{\mu}$, invariant.  $BMS$  group is smaller than   $Diff(\mathcal{I})$ and has an amazing structure, as it does not change normal vectors of $\mathcal{I}$. This causes the relation
      	 \begin{equation}
      	\mathcal{L}_{\xi}\tilde{n}^{\mu}|_{\mathcal{I}}=\alpha \tilde{n}^{\mu},
      	\end{equation} 
      	where  $\xi^{\mu}$ is the $BMS$ vector field and  $\alpha$ is function that satisfies $\mathcal{L}_{n}\alpha|_{\mathcal{I}}=0$.
      	 $BMS$ translations have to preserve  $\tilde{n}_{\mu}\tilde{n}^{\mu}$ on  ${\mathcal{I}}$. To have more senses about The $BMS$ group, it would be useful to consider the intrinsic metric (see Appendix \ref{app:A})
      	 \begin{equation}
      	 ds^2=d\xi d\xi^*=-1/4(1+\xi \xi^*)^2(d\theta^2+\sin^2\theta d\varphi^2),
      	 \end{equation}
      	 where $\xi=e^{i\varphi}cot{\theta/2}$. If one chooses the conformal factor $\Omega=\frac{2}{(1+\xi \xi^*)}$ each cut would be a 2-sphere. This coordinate system is useful to find the symmetry group. For a sphere the holomorphic bijections have the form \cite{esposito1992mathematical}
      	 \begin{equation}
      	 f(\xi)=\frac{a\xi+b}{c\xi+d},
      	 \label{eq:4.777}
      	 \end{equation}
      	 where $ad-bc=1$. \eqref{eq:4.777} transformations are known as fractional linear transformations. The following  conformal transformations would be valid if \eqref{eq:4.777} transformations preserve the intrinsic metric of each cut.
      	 \begin{equation}
      	 d\Sigma'^2=\omega^2d\Sigma^2\quad,\quad d\Sigma^2=d\theta^2+\sin^2\theta d\varphi^2.
      	 \end{equation}
      	 For $(q_{\mu\nu},n^{\mu})$ one can write
      	 \begin{equation}
      	 (q_{\mu\nu},n^{\mu}) \rightarrow (\omega^{2}q_{\mu\nu},\omega^{-1}n^{\mu}).
      	 \end{equation}
      	 Thus it is possible to find the conformal factor $\omega$ by calculating $d\Sigma'$
      	 \begin{align}
      	 \label{eq:4.1222}
      	 dS'&=d\xi'd\xi'^*\\
      	 &=\frac{ad\xi(c\xi+d)-cd\xi(a\xi+b)\times(a^*d\xi^*(c^*\xi^*+d^*)-c^*d\xi^*(a^*\xi^*+b^*)}{(c\xi+d)^2(c^*\xi^*+d^*)^2}\notag\\
      	 &=\frac{\overbrace{ad}^{bc+1}a^*d^*+\overbrace{cb}^{ad-1}c^*b^*-ca^*bd^*-ac^*db^*}{(c\xi+d)^2(c^*\xi^*+d^*)^2}d\xi d\xi^*\notag\\
      	 &=\frac{d\xi d\xi^*}{(c\xi+d)^2(c^*\xi^*+d^*)^2}=\frac{-(1+\xi \xi^*)d\Sigma^2}{4(c\xi+d)^2(c^*\xi^*+d^*)^2}.\notag
      	 \end{align} 
      	 On the other hand
      	 \begin{align}
      	 \label{eq:4.1333}
      	 dS'&=-\frac{1}{4}(1+\xi'\xi'^*)^2d\Sigma'^2\\
      	 &=(\frac{-(a^*\xi^*+b^*)(a\xi+b)+(c\xi+d)(c^*\xi^*+d^*)}{4(c\xi+d)(c^*\xi^*+d^*)})d\Sigma'^2\notag.
      	 \end{align}
      	 The equation
      	 \eqref{eq:4.1222}
      is equal to
      	 \eqref{eq:4.1333}
      	 thus
      	 	 \begin{equation}
      	 d\Sigma'^2=\frac{(1+\xi\xi^*)^2}{[(a\xi+b)(a^*\xi^*+b^*)+(c\xi+d)(c^*\xi^*+d^*)]^2}d\Sigma^2,
       \end{equation}
       so  $\omega$ is
       \begin{equation}
       \omega=\frac{1+\xi\xi^*}{(a\xi+b)(a^*\xi^*+b^*)+(c\xi+d)(c^*\xi^*+d^*)},
       \end{equation}
       where $\mathcal{L}_{n}\omega=0$ and the line element, in the direction of $\mathcal{I}$ generators, changes as
       \begin{equation}
       du'=\omega du \quad \rightarrow\quad u'=\omega[u+\alpha(\xi,\xi^*)].
       \label{eq:4.111}
       \end{equation}
       \eqref{eq:4.777} till \eqref{eq:4.111} are transformations from the $BMS$ group \cite{boyle2016transformations}.
       \subsection{supertranslations}
 	     If the component $u$ in the direction of $\mathcal{I}$ generators transforms as 
 	     \begin{equation}
 	     \hat{u}=u+\alpha(\xi,\xi^*),
 	     \label{eq:4.1111}
 	     \end{equation}
 	     This is a supertranslation.   In 1966 Newman and Penrose proposed to write $\alpha$ in
 	     terms of spherical harmonics \cite{sachs1962asymptotic,newman1966note}
 	     \begin{equation}
 	     \alpha=\Sigma_{l=0}^{\infty}\Sigma_{m=-l}^{l}a_{l,m}Y_{l,m}(\theta,\varphi)
 	     \end{equation}
 	     where $a_{l,m}$ is constant. If $a_{l,m}=0$ for $ l>2$ then 
 	     \begin{equation}
 	     \alpha=\epsilon_0+\epsilon_1 \sin \theta \cos \varphi +\epsilon_2 \sin \theta \sin \varphi+ \epsilon_3 \cos theta,
 	     \end{equation}
 	     Here the supertranslations reduce to a special case,
 	     called the translations.\cite{newman1966note}
 	     \subsection{translations}
 	     Translations in Minkowski space-time can be written as
 	     \begin{equation}
 	     \label{eq:4.122}
 	     t'=t+a\quad,\quad  x'=x+b\quad,\quad  y'=y+c\quad,\quad  z'=z+d.
 	     \end{equation}
 	     One can define a coordinate system as
 	     \begin{align}
 	     \label{eq:4.133}
 	     &u=t-r,\\
 	     &r^2=x^2+y^2+z^2,\notag\\
 	     &\xi=e^{i\varphi}\cot{\theta/2},\notag\\
 	     &Z=\frac{1}{1+\xi\xi^*}.\notag
 	     \end{align}
 	     It is possible to write $x$, $y$ and $z$ according to complete conjugate variables as 
 	     \begin{equation}
 	     \label{eq:4.21}
 	     x=r(\xi+\xi^*)Z\quad,\quad y=-ir(\xi-\xi^*)Z\quad,\quad
 	     z=r(\xi\xi^*-1)Z.
 	     \end{equation}
 	     Using \eqref{eq:4.122},
 	     \eqref{eq:4.133}
 	     and
 	     \eqref{eq:4.21}
 	     $r'$ can be obtained as
 	     \begin{align}
 	     r'&=\sqrt{x'^2+y'^2+z'^2}\\
 	     &=(r^2(\xi+\xi^*)^2Z^2+b^2+2r(\xi+\xi^*)Zb-r^2(\xi-\xi^*)^2Z^2-c^2+ir(\xi-\xi^*)Zc\notag\\
 	     &+r^2(\xi\xi^*-1)^2Z^2+d^2+2r(\xi\xi^*-1)Zd)^{1/2}\notag\\
 	     &=(r^2Z^2\underbrace{(\xi^2\xi^{*2}+2\xi\xi^*+1)}_{(\xi\xi^*+1)^2}+2rZ[\underbrace{(b+ic)}_{B}\xi+\underbrace{(b-ic)}_{B^*}\xi^*+2rZ(\xi\xi^*d-d)]+c')1/2\notag\\
 	     &=rZ(\xi\xi^*+1)\sqrt{1+\frac{\frac{2}{rZ}(B\xi+B^*\xi^*+\xi\xi^*d-d)+\frac{c'}{r^2Z^2}}{(\xi\xi^*+1)^2}}\notag\\
 	     &\Rightarrow\notag\\
 	     r'&\simeq(\xi\xi^*+1)[rZ+\frac{B\xi+B^*\xi^*+\xi\xi^*d-d}{(\xi\xi^*+1)^2}+O(1/r)]\notag\\
 	     &=r+\frac{B\xi+B^*\xi^*+\xi\xi^*d-d}{(\xi\xi^*+1)}+O(1/r)\notag.
 	     \end{align}
 	     Thus for $u'=t'-r'$, it is possible to write
 	     \begin{align}
 	     u'&=t'-r'=t+a-r-\frac{B\xi+B^*\xi^*+\xi\xi^*d-d}{(\xi\xi^*+1)}+O(1/r)\\
 	     &=u-\frac{B\xi+B^*\xi^*+\overbrace{(-a+d)}^{C}\xi\xi^*+\overbrace{(-a-d)}^{A}}{(\xi\xi^*+1)}+O(1/r)\notag.
 	     \end{align}
 	     Therefore \begin{equation}
 	     u'=u+({A+B\xi+B^*\xi^*+C\xi\xi^*})Z+O(1/r).
 	     \end{equation}
 	     Thus if someone put the relation
 	     \begin{equation}
 	     \alpha=\frac{A+B\xi+B^*\xi^*+C\xi\xi^*}{1+\xi\xi^*},
 	     \label{eq:12}
 	     \end{equation}
 	     in \eqref{eq:4.1111} supertranslations will be obtained. So the asymptotic symmetry group for such a space-time is the subgroup of $Diff(\mathcal{I})$ that preserve the fall of the intrinsic metric, $q_{\mu\nu}$ that means the fall of $\Omega$ and its derivatives.This group is the $BMS$ group. As a comparison Pioncaré group, $\mathfrak{P}$, is obtained by the semidirect product of the translations group,
 	     $\mathfrak{T}$, to Lorentz group, $\mathfrak{L}$ \cite{barnich2014notes},
 	     \begin{equation}
 	     \mathfrak{P}=\mathfrak{L}\rtimes \mathfrak{T}.
 	     \end{equation}
 	     Thus the $BMS$ group will obtained as follows
 	     \begin{equation}
 	     \mathfrak{B}=\mathfrak{L}\rtimes \mathfrak{S},
 	     \end{equation}
       which the four dimensional translations group,
$\mathfrak{T}$, is replaced with the infinite dimensional supertranslations, $\mathfrak{S}$, this generators are $fn^{\mu}$ vector fields on $\mathcal{I}$ where $f$ is a scalar variable that satisfies $\mathcal{L}_{n}f=0$. In other words, the $BMS$ group is a group that maps   $\mathcal{I}^{+}$   on itself. 
  
 	     \section{Asymptotic fields}
       A great motif to find asymptotic symmetries is the problem of defining conserved charges in gauge theories, like electric charges in electrodynamics and energy momentum in general relativity. This problem is a result of the Noether-charge puzzle for gauge symmetries. In fact, the problem is that when one tries to define a conserved charge according to Noether's first theorem, Noether's current vanishes on shell. To be more explicit, one can consider a scalar field, $\varphi^{i}$ and the Lagrangian, $L[\varphi]$. The Euler-Lagrange equation is \cite{compere2019advanced} 
 	     \begin{equation}
 	     	\frac{\delta L}{\delta \Phi^{i}}= \frac{\partial L}{\partial \Phi^{i}}-\partial_{\mu}\left(\frac{\partial L}{\partial \partial_{\mu} \Phi^{i}}\right)+\partial_{\mu} \partial_{v}\left(\frac{\partial L}{\partial \partial_{\mu} \partial_{v} \Phi^{i}}\right)+\cdots
 	     \end{equation}
 	     where $\forall \Phi^{i} \in \Phi=\left\{\left(\Phi_{M}^{i}\right)_{i \in I}, g_{\mu v}\right\}$.
 	     The generators for this system are shown as below  \cite{cotaescu2000external}
 	     \begin{equation}
 	     	\delta_{f} \phi^{i}=R^{i}_{\alpha}(f^{\alpha})=R^{i}_{\alpha},
 	     \end{equation}
 	     where $f^{\alpha}$ is an arbitrary function and satisfies the following relation
 	     \begin{equation}
 	     	\label{eq:6.1}
 	     	R^{i}_{\alpha}(f^{\alpha}) \frac{\delta L}{\delta \phi^{i}} =\partial _{\mu} j^{\mu}_{f} .
 	     \end{equation}
 	     In consonance with Noether's second theorem, one has
 	     \begin{equation}
 	     	R^{+i}_{\alpha}\frac{\delta L}{\delta \phi^{i}}=0,
 	     \end{equation}
 	     where $R^{+i}_{\alpha}$ is written without concerning the boundary terms. $R^{+i}_{\alpha}$ has  the following relation with local operator, $Q_{i}$ 
 	     \begin{equation}
 	     	R^{+i}_{\alpha}=\sum_{k=0}^{}i^{k} \partial_{\mu_{1}} \dots \partial_{\mu_{k}} [R^{i(\mu_{1} \dots \mu_{k})}]_{\alpha} Q_{i}.
 	     \end{equation}
 	     In presence of boundary terms one can write
 	     \begin{equation}
 	     	\forall Q_{i}, f^{\alpha}: Q_{i}R^{i}_{\alpha}(f^{\alpha})=f^{\alpha}R^{+i}_{\alpha}(Q_{i})+\partial_{\mu}S^{\mu i}_{\alpha}(Q_{i},f^{\alpha}) , 
 	     \end{equation}
 	     where $S^{\mu i}_{\alpha}$ are differential equations.
 	     If
 	     $Q_{i}=\frac{\delta L}{\delta \varphi^{i}}$,  then \cite{barnich2008surface}
 	     \begin{equation}
 	     	\frac{\delta L}{\delta \varphi^{i}}R^{i}_{\alpha}(f^{\alpha})= \partial_{\mu} S^{\mu i}_{\alpha}(\frac{\delta L}{\delta \varphi^{i}},f^{\alpha}) \label{eq:6.7} 
 	     \end{equation} 
 	     Thus $S^{\mu i}_{\alpha}(\frac{\delta L}{\delta \varphi^{i}},f^{\alpha})$  is the Noether's current that satisfies the equation \eqref{eq:6.1}. This current vanishes on shell because of the linear distribution of  $(\frac{\delta L}{\delta \varphi^{i}},f^{\alpha})$. According to \eqref{eq:6.1} and  \eqref{eq:6.7}, it is possible to write 
 	     \begin{equation}
 	     	\partial_{\mu}(j_{\pm}^{\mu}-S^{\mu i}_{\alpha}(\frac{\delta L}{\delta \varphi^{i}},f^{\alpha}))=0.
 	     \end{equation}
 	     Considering Poincaré lemma, one can obtain the following relation on shell
 	     \begin{equation}
 	     	j_{\pm}^{\mu}=S^{\mu i}_{\alpha}(\frac{\delta L}{\delta \varphi^{i}},f^{\alpha})- \partial_{\nu}k_{f}^{[\nu \mu]}\quad,\quad n>1 
 	     \end{equation}
 	     where $k_{f}^{[\nu \mu]}$  is the superpotential and $n$ is the space-time's dimension. In one dimension, it is possible to write \cite{barnich2002covariant}
 	     \begin{equation}
 	     	j_{f}=S^{\mu i}_{\alpha}(\frac{\delta L}{\delta \varphi^{i}},f^{\alpha})+C,
 	     \end{equation} 
 	     where  $C$ is an arbitrary constant. This relation is the solution of \eqref{eq:6.1}.  $k_{f}^{[\nu \mu]}$ is arbitrary  because according to  $\partial_{\nu} \partial_{\mu} k^{[\nu \mu]}_{f}=0$, it vanishes in  \eqref{eq:6.1}. This means Noether's currents are undefinable. In the case of exact solutions and symmetries, surface charges in the full theory are
 	     constructed by integrating the surface charge 1-forms of the linearized theory along a
 	     path in the space of symmetric configurations \cite{barnich2008surface}
 	     \begin{equation}
 	     	n>1: Q[\varphi (x)]= \int_{\Sigma} j_{f} |_{\varphi (x)}= \int_{\partial \Sigma} k_{f} |_{\varphi (x)}, \label{eq:(11.6)}
 	     \end{equation}
 	     which is the solution of the Euler-Lagrange equation. In this relation, $\Sigma$  is a $n-1$ dimensional spacelike surface with the boundary $\partial \Sigma$,  $n-1$-form current $j_{f}$ and $n-2$-form current  $k_{f}$ that is the superpotential
 	     \begin{align}
 	     	&j_{f}=j_{f}^{\mu}(d^{n-1}x)_{\mu},\\
 	     	&k_{f}=k_{f}^{[\mu \nu]}(d^{n-2}x)_{\mu \nu},\notag\\
 	     	&(d^{n-p}x)_{\mu_{1}\dots \mu_{p}} := \frac{1}{p! (n-p)!} \epsilon_{\mu_{1}\dots \mu_{p}} dx^{\mu_{(p+1)}} \dots dx^{\mu_{n}}. \notag
 	     \end{align} 
      As the relation \eqref{eq:(11.6)}  shows the mentioned problem, it also  proposes a way to solve it. Actually, the relation \eqref{eq:(11.6)}  is dependent on the boundary's features of superpotentials similar to the electrodynamics where  $F_{\mu\nu}$ is written in agreement with to its superpotentials. So it may be possible to do the same thing here that proposes  a probable relation  between gauge symmetries and $(n-2)$-forms. The superpotential has been obtained by Abbott and Deser for asymptotically flat space-time \cite{abbott1982stability}
 	     \begin{equation}
 	     	\mathbf{k}_{\zeta}^{\mu v}[h ; g]=\frac{\sqrt{-g}}{8 \pi G}\left(\xi^{\mu} \nabla_{\sigma} h^{v \sigma}-\xi^{\mu} \nabla^{v} h+\xi_{\sigma} \nabla^{v} h^{\mu \sigma}+\frac{1}{2} h \nabla^{v} \xi^{\mu}-\frac{1}{2} h^{\rho \nu} \nabla_{\rho} \xi^{\mu}+\frac{1}{2} h_{\sigma}^{v} \nabla^{\mu} \xi^{\sigma}\right).
 	     \end{equation} 
 	      \section{De Sitter space-time}
 	      \checkmark First definition: If  $\tilde{M}$ with $\mathcal{I}$ as its boundary and with a diffeomorphism between ${M}$    $(\tilde{M}\setminus \mathcal{I})$ exists, then $({M},{g}_{\mu\nu})$  is weakly asymptotically de Sitter as \cite{ashtekar2014asymptotics}
 	      
 	      1. A smooth function  $\Omega$ exists on  $\tilde{M}$ that vanishes on $\mathcal{I}$ and  $n_{\mu}:=\tilde{\nabla}_{\mu}\Omega|_{\mathcal{I}}\neq0$. In addition one may be able to write  $ \tilde{g}_{\mu\nu}=\Omega^{2} {g}_{\mu\nu}$. 
 	      
 	      2. ${g}_{\mu\nu}$ must satisfy the Einstein equation in presence of a positive cosmological constant \cite{ashtekar2014asymptotics}
 	      \begin{equation}
 	      {R}_{\mu\nu}-1/2{R}{g}_{\mu\nu}+\Lambda {g}_{\mu\nu}=8 \pi G {T}_{\mu\nu}\quad , \quad \Lambda>0.
 	      \end{equation} 
 	      These two conditions are similar to the conditions defined for asymptotically flat space-time. The first condition shows the relation between the physical and unphysical space-times and  $\nabla_{\mu} \Omega \neq 0$ assure one to be able to use $\Omega$ as a component on   $\mathcal{I}$. The second condition declares that $\Omega^{-2}T_{\mu\nu}$ decrease smoothly on $\mathcal{I}$.
 	      Fortunately it is possible to have different choices for $\Omega$ and one can see in the following sections how this feature help us.
 	      
\begin{figure}
 	      	\centering
 	      	\begin{tikzpicture}[scale=3]
 	      		\draw (0,0) ellipse (0.75 and 0.3);
 	      		\draw (-0.75,0) -- (-0.75,-2);
 	      		\draw (-0.75,-2) arc (180:360:0.75 and 0.3);
 	      		\draw [dashed] (-0.75,-2) arc (180:360:0.75 and -0.3);
 	      		\draw (0.75,-2) -- (0.75,0);  
 	      		\fill [yellow!40,opacity=0.5] (-0.75,0) -- (-0.75,-2) arc (180:360:0.75 and 0.3) -- (0.75,0) arc (0:180:0.75 and -0.3);
 	      		\node at (0,-0.2){$\mathcal{I}^+$};
 	      		\node at (0,-2.2){$\mathcal{I}^-$};
 	      	\end{tikzpicture}
 	      	\caption{ In this figure the upper circle is $\mathcal{I}^+$ and the lower circle is $\mathcal{I}^-$. The topology of this two circles are $\mathbb{S}^{3}$. \label{pic:4.1}}
 	      \end{figure}
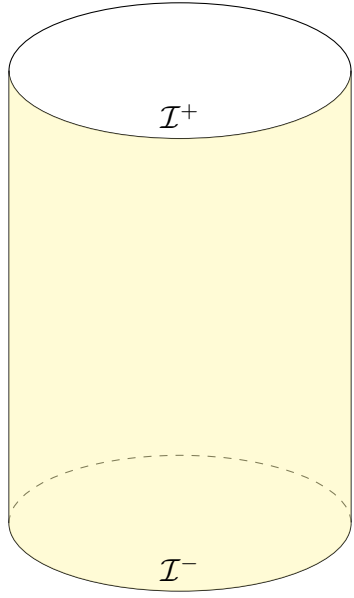

 	      \checkmark Second definition: if  $\mathcal{I}$ was spacelike and geodesically complete then  a weakly asymptotically de Sitter space-time will be a asymptotically de Sitter space-time. It is not possible to distinguish the topology of $\mathcal{I}$ but three possible topology have been studied by Ashtekar et al \cite{ashtekar2014asymptotics}:
 	      
 	      $\bullet$ If   $\mathcal{I}$ has the topology   $\mathbb{S}^{3}$ then the space-time will be known as globally asymptotically  de  Sitter (figure \ref{pic:4.1}).
 	      
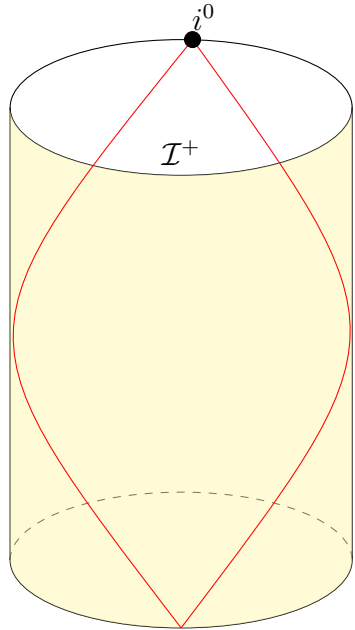
\begin{figure}
 	        	\centering
 	        	\begin{tikzpicture}[scale=3]
 	        		\draw (0,0) ellipse (0.75 and 0.3);
 	        		\draw (-0.75,0) -- (-0.75,-2);
 	        		\draw (-0.75,-2) arc (180:360:0.75 and 0.3);
 	        		\draw [dashed] (-0.75,-2) arc (180:360:0.75 and -0.3);
 	        		\draw (0.75,-2) -- (0.75,0);  
 	        		\fill [yellow!40,opacity=0.5] (-0.75,0) -- (-0.75,-2) arc (180:360:0.75 and 0.3) -- (0.75,0) arc (0:180:0.75 and -0.3);
 	        		\draw[red]  (0, -2.3) .. controls (0.98,-1) .. (0.05,0.3);
 	        			\draw[red]  (0, -2.3) .. controls (-0.99,-1) .. (0.05,0.3);
 	        		\node at (0,-0.2){$\mathcal{I}^+$};
 	        		\node at (0.1,0.4){$i^0$};
 	        		\filldraw (0.05,0.3) circle[radius=1pt];
 	        	\end{tikzpicture}
 	        	\caption{ The topology of null infinity in space-time that is asymptotically de Sitter in Pioncaré patch is $\mathbb{R}^{3}$.\label{pic:4.2}}
 	        \end{figure}

 	        $\bullet$ 
 	        $\mathcal{I}$ with the topology  $ \mathbb{R}^{3} \simeq \mathbb{S}^{3} \setminus \left \{ p \right \}$ results a space-time that is asymptotically de Sitter in Pioncaré patch, where  $p$ is spacelike infinity,  $i^0$ (figure \ref{pic:4.2}).

\begin{figure}
 	        	\centering
 	        	\begin{tikzpicture}[scale=3]
 	        		\draw (0,0) ellipse (0.75 and 0.3);
 	        		\draw (-0.75,0) -- (-0.75,-2);
 	        		\draw (-0.75,-2) arc (180:360:0.75 and 0.3);
 	        		\draw [dashed] (-0.75,-2) arc (180:360:0.75 and -0.3);
 	        		\draw (0.75,-2) -- (0.75,0);  
 	        		\fill [yellow!40,opacity=0.5] (-0.75,0) -- (-0.75,-2) arc (180:360:0.75 and 0.3) -- (0.75,0) arc (0:180:0.75 and -0.3);
 	        		\draw [dashed](0.1,0.3) -- (0.1,-1.7);
 	        		\draw[red] (0.05,- 0.3) .. controls (0.955,-0.5) .. (0.1, -1.3) ;
 	        		\draw[red] (-0.25,- 0.3) .. controls (-0.95,-0.5) .. (0.1, -1.3) ;
 	        		\draw[red]  (-0.25, -2.3) .. controls (-0.95,-1.5) .. (0.1, -1.3);
 	        		\draw[red]  (0.05, -2.3) .. controls (0.955,-1.5) .. (0.1, -1.3);
 	        		\draw[yshift = 0.1cm, snake] (0.05,- 0.38) -- (-0.25,- 0.38);
 	        		\draw[yshift = 0.1cm, snake] (0.05,- 2.38) -- (-0.25,- 2.38);
 	        		\node at (0.55,0.3){$\mathcal{I}^+$};
 	        		\node[gray] at (-0.1,-0.2){$r=0$};
 	        		\node at (0.1,0.4){$i^0$};
 	        			\node at (0.2,- 0.2){$i^+$};
 	        		\node at (-0.35,- 0.2){$i^-$};
 	        		\filldraw (0.05,0.3) circle[radius=1pt];
 	        		\filldraw (0.05,- 0.3) circle[radius=1pt];
 	        		\filldraw (-0.25,- 0.3) circle[radius=1pt];
 	        		\filldraw (0.05,- 2.3) circle[radius=1pt];
 	        		\filldraw (-0.25,- 2.3) circle[radius=1pt];
 	        		\draw [blue](0.05,- 2.3) --(-0.25,- 0.3);
 	        		\draw [blue] (0.05,- 0.3) -- (-0.25,- 2.3);
 	        	\end{tikzpicture}
 	        	\caption{This figure shows the dS-Schwarzschild space-time where red lines are dS horizons and blue line are event horizons. The topology of null infinity is  $\mathbb{R} \times \mathbb{S}^{2}$.  \label{pic:4.3}}
 	        \end{figure}
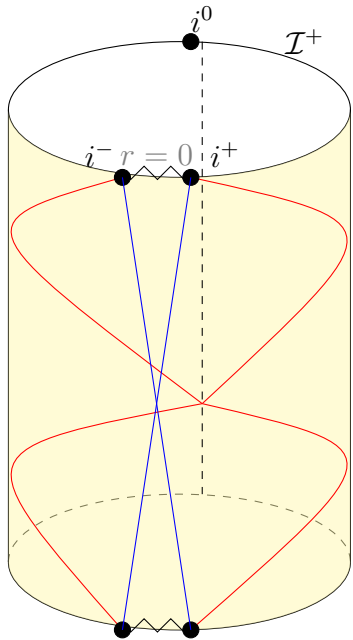

 	        $\bullet$  The space-time,  $({M},{g}_{\mu\nu})$ is asymptotically de Sitter-Shcwartzshild if  $\mathcal{I}$ has the topology    $\mathbb{R} \times \mathbb{S}^{2} \simeq \mathbb{S}^{3}\setminus \left \{ p_1 ,p_2 \right \}$, where    $p_1$ is  $i^{\pm} $ and    $p_2$ is    $i^0$ (figure \ref{pic:4.3}).
 	        
 	        Definitely all of these definitions have to satisfy the condition one. These points that present infinity make trouble for further calculations. So it would be useful to work with the first topology.
 	        
 	          \checkmark Third definition:
            $({M},{g}_{\mu\nu})$ is strongly asymptotically de Sitter if it satisfies the second definition and the intrinsic metric,  $q_{\mu\nu}$ becomes conformally flat \cite{ashtekar2014asymptotics}.
 	          
 	          These definitions have been written according to the choice of conformal factor thus each space-time can be included to different groups. Generally the metric be written as follows near the infinity
 	          \begin{equation}
 	          \tilde{g}_{\mu\nu}=-\tilde{\nabla}_{\mu}\Omega\tilde{\nabla}_{\nu}\Omega+\tilde{h}_{\mu\nu}
 	          \label{eq:4.metric}
 	          \end{equation}
 	          where $\tilde{h}_{\mu\nu}$ is a function of $\Omega$.
 	          \section{Asymptotic de Sitter space-time's symmetries}
 	          As intrinsic metric in 
 	          $\Lambda>0$ case takes the signature $(+,+,+)$, $n^{\mu}$ is no longer tangential to $\mathcal{I}$. Thus it is not possible to follow the same route as in $\Lambda=0$ case to find the asymptotic symmetry group. A good idea is to consider the intrinsic metric to be  conformally flat  then the symmetry group reduces to   $So(4,1)$. Restricting the conditions, Bach tensor vanishes and this causes losing information by fiat. In this section we review the work of Ashtekar et. al. \cite{ashtekar2014asymptotics}.
 	          
 	          So finding another framework is essential. To achieve this aim, Einstein equations has been rewritten according to the conformal rescaling $\Omega$ \cite{ashtekar2014asymptotics}
 	          \begin{equation}
 	          \tilde{R}_{\mu\nu}-1/2 \tilde{g}_{\mu\nu}\tilde{R}+ 2 \Omega^{-1}(\tilde{\nabla}_{\mu}n_{\nu}-\tilde{g}_{\mu\nu}\tilde{\nabla}^{\sigma}n_{\sigma})+3\Omega^{-2} \tilde{g}_{\mu\nu}n^{\sigma}n_{\sigma}+\Omega^{-2} \Lambda \tilde{g}_{\mu\nu}=8 \pi G{T}_{\mu\nu}, \label{eq :4.5}
 	          \end{equation}
 	          where $\tilde{n}_{\mu}:=\tilde{\nabla}_{\mu}\Omega$ (see Appendix \ref{ap:d}).
 	          
 	         If one multiplies $\Omega^{2}$  by the relation \eqref{eq :4.5} and applies boundary conditions, which has been mentioned in the first definition, in the resulting equation, then
 	         
 	         \begin{equation}
 	         \tilde{n}^{\mu}\tilde{n}_{\mu}|_{\mathcal{I}}= - \Lambda/3 =-1/l^{2}.
 	         \label{eq:4.39}
 	         \end{equation}
 	         Thus 
 	         $\tilde{n}_{\mu}$ is timelike on $\mathcal{I}$ and as a result $\mathcal{I}$ itself is spacelike. Because of the exciting freedom to choose the conformal factor,
 	         it is possible to choose one that satisfy the relation
 	          $\tilde{\nabla}_{\mu}\tilde{n}^{\mu} |_{\mathcal{I}}=0$. This choice considerably  simplifies  mathematical calculations. Now multiplying $\Omega$ by \eqref{eq :4.5} and considering \eqref{eq:4.39}, the third and fourth terms of \eqref{eq :4.5}  simplify together, thus
 	         \begin{equation}
 	         \label{eq:4.65}
 	         \tilde{\nabla}_{\mu}\tilde{n}_{\nu}|_{\mathcal{I}}=0.
 	         \end{equation}
 	         Applying all these restriction one conformal degree of freedom is left, $\Omega \rightarrow \Omega'=\omega \Omega$ that results $\tilde{n}^{\mu}\tilde{\nabla}_{\mu}\omega |_{\mathcal{I}}=0$. Using this conformal frame, $\tilde{C}_{\mu\nu \sigma \lambda}$ vanishes near $\mathcal{I}$. To show this, Schouten tensor is represented in the following \cite{hollands2005comparison}
 	         \begin{equation}
 	         \tilde{S}_{\mu\nu}:=\tilde{R}_{\mu\nu}-(\tilde{R}/6)\tilde{g}_{\mu\nu}.
 	         \end{equation}
 	         The relation between Schouten tensor and its conformal transformed form is (see Appendix \ref{ap:d}) 
 	         \begin{equation}
 	         \tilde{S}_{\mu\nu}=S_{\mu\nu}-2 \Omega^{-1}\tilde{\nabla}_{\mu}\tilde{n}_{\nu}+2\Omega^{-2}\tilde{g}_{\mu\nu}\tilde{n}^{\sigma}\tilde{n}_{\sigma}. \label{eq:8.5}
 	         \end{equation}
 	         On the other hand it is possible to write conformal transformed Riemann tensor (see Appendix \ref{ap:j})
 	         \begin{equation}
 	         \tilde{R}_{\mu\nu \sigma \lambda}=\tilde{C}_{\mu\nu \sigma \lambda}+\tilde{g}_{\mu[\sigma}\tilde{S}_{\lambda]\nu}-\tilde{g}_{\nu[\sigma}\tilde{S}_{\lambda]\mu}, \label{eq:5.9}
 	         \end{equation}
 	         multiplying this relation  by $\tilde{n}^{\lambda}$
 	          \begin{align}
 	         \label{eq:4.nR}
 	         &\underbrace{\tilde{n}^{\lambda}\tilde{R}_{\mu\nu \sigma \lambda}}_{=0}=\tilde{n}^{\lambda}\tilde{C}_{\mu\nu \sigma \lambda}+\tilde{g}_{\sigma[\mu}\tilde{S}_{\nu]\lambda}\tilde{n}^{\lambda}-\tilde{n}_{[\mu}\tilde{S}_{\nu]\sigma}\\
 	         &\Rightarrow \tilde{n}_{[\mu}\tilde{S}_{\nu]\sigma}=\tilde{n}^{\lambda}\tilde{C}_{\mu\nu \sigma \lambda}+\tilde{g}_{\sigma[\mu}\tilde{S}_{\nu]\lambda}\tilde{n}^{\lambda}\notag.
 	         \end{align} 
 	         where $\tilde{n}^{\lambda}\tilde{R}_{\mu\nu \sigma \lambda}=0$ has been used. On the other hand  multiplying \eqref{eq:8.5} by $\Omega$ and taking the derivative, one has
 	         
 	         \begin{align}
 	         \label{eq:5.11}
 	         \tilde{\nabla}_{[\mu}\Omega\tilde{S}_{\nu]\sigma}&=\Omega \tilde{\nabla}_{[\mu}\tilde{S}_{\nu]\sigma}+\tilde{n}_{[\mu} \tilde{S}_{\nu]\sigma}=\Omega \tilde{\nabla}_{[\mu}\tilde{S}_{\nu]\sigma}+
 	         \tilde{n}^{\lambda}\tilde{C}_{\mu\nu \sigma \lambda}+\tilde{g}_{\sigma[\mu}\tilde{S}_{\nu]\lambda}\tilde{n}^{\lambda}\\
 	         &=\Omega \tilde{\nabla}_{[\mu}{S}_{\nu]\sigma}+4(\tilde{\nabla}_{[\mu}\Omega )\Omega^{-2}\tilde{g}_{\nu]\sigma}\tilde{n}_{\lambda}\tilde{n}^{\lambda}+
 	         \tilde{n}^{\lambda}\tilde{C}_{\mu\nu \sigma \lambda}+\tilde{g}_{\sigma[\mu}{S}_{\nu]\lambda}\tilde{n}^{\lambda}\notag\\
 	         &-2\Omega^{-2}\tilde{n}_{[\mu}(\tilde{\nabla}_{\nu]}\tilde{n}_{\lambda}+2\Omega^{-2}\tilde{n}_{[\mu}(\tilde{\nabla}_{\nu]}\tilde{n}_{\lambda}-4(\tilde{\nabla}_{[\mu}\Omega )\Omega^{-2}\tilde{g}_{\nu]\sigma}\tilde{n}_{\lambda}\tilde{n}^{\lambda}
 	         \notag\\
 	         &-2\tilde{\nabla}_{[\mu}\tilde{\nabla}_{\nu]}\tilde{n}_{\sigma}=\Omega \tilde{\nabla}_{[\mu}{S}_{\nu]\sigma}+\tilde{n}^{\lambda}\tilde{g}_{\lambda[\mu} S_{\nu]\sigma}+\tilde{C}_{\mu\nu\sigma\lambda}\tilde{n}^{\lambda},\notag
 	         \end{align}
 	         where  $\tilde{K}_{\mu\nu}\tilde{n}^{\nu}=0$ and  \eqref{eq:4.nR} has been used. Additionally for fields' equations in physical space-time one can write \cite{ashtekar2014asymptotics}
 	         \begin{equation}
 	         {S}_{\mu\nu}=(\Lambda/3){g}_{\mu\nu}+8 \pi G({T}_{\mu\nu}-1/3 {T}
 	         {g}_{\mu\nu}) \equiv \Lambda /3 {g}_{\mu\nu}+\bar{T}_{\mu\nu},
 	         \end{equation}
 	         where $\bar{T}_{\mu\nu}:=8 \pi G({T}_{\mu\nu}-1/3 {T}
 	         {g}_{\mu\nu})$. This relation can be used in \eqref{eq:5.11} so
 	         \begin{equation}
 	         \Omega \tilde{\nabla}_{[\mu}{S}_{\nu]\sigma}+\tilde{C}_{\mu\nu\sigma\lambda}n^{\lambda}=\tilde{\nabla}_{[\mu}(\Omega\bar{T}_{\nu]\sigma})-g_{\sigma[\mu}\bar{T}_{\nu]\lambda}n^{\lambda}.
 	         \label{eq:5.13}
 	         \end{equation}
 	         As $\Omega^{-1
 	         } {T}_{\mu\nu}$ has a smooth limit on  $\mathcal{I}$, it is possible to say
          \begin{equation}
          \tilde{C}_{\mu\nu\sigma\lambda}n^{\lambda}|_{\mathcal{I}}=0.
          \label{eq:4.722}
          \end{equation}
          Weyl tensor can be devided to electric an magnetic parts
          \begin{align}
          \label{eq:5.24}
          &\tilde{E}_{\mu\sigma}:=l^{2}\tilde{C}_{\mu\nu\sigma\lambda}\tilde{n}^{\nu}\tilde{n}^{\sigma},\\
          &\tilde{B}_{\mu\sigma}:=l^{2} \star \tilde{C}_{\mu\nu\sigma\lambda}\tilde{n}^{\nu}\tilde{n}^{\sigma},\notag
          \end{align}
          where 
           \begin{equation}
          \star C_{\mu\nu\sigma\rho}\equiv C_{\mu\nu\sigma\rho}+iC_{\mu\nu\sigma\rho}^{\sim},
          \end{equation}
          for more details see Appendix \ref{ap:j}. In this relation $C_{\mu\nu\sigma\rho}^{\sim}$ is the right dual. both relations in \eqref{eq:5.24}  vanish on $\mathcal{I}$ because of \eqref{eq:4.722} so
          \begin{equation}
          \label{eq:4.75}
          \tilde{C}_{\mu\nu\sigma\lambda}|_{\mathcal{I}}=0.
          \end{equation}
          
          \section{Asymptomatic expansion}
          As it is shown in the previous  section, considering the intrinsic metric to be conformally flat is not a good method to define the asymptotic symmetry group as it causes losing information by fiat. So it is necessary to find another method. Here the Fefferman-Graham framework is presented. As it said before, the metric near $\mathcal{I}$ can be defined as \eqref{eq:4.metric}. To find  $\tilde{h}_{\mu\nu}$, one can start with its Lie derivative, \cite{fefferman1985conformal} $\tilde{h}_{\mu\nu}$
          \begin{equation}
          \mathcal{L}_{n}\tilde{h}_{\mu\nu}=\underbrace{\tilde{n}^{\sigma}\tilde{\nabla}_{\sigma}\tilde{h}_{\mu\nu}}_{=0}-\tilde{h}_{\mu\sigma}\underbrace{\tilde{\nabla}_{\nu}\tilde{n}^{\sigma}}_{\tilde{K}^{\sigma}_{\nu}}-\tilde{h}_{\nu\sigma}\underbrace{\tilde{\nabla}_{\mu}\tilde{n}^{\sigma}}_{\tilde{K}^{\sigma}_{\mu}}=-2{K}_{\mu\nu}
          \end{equation}
          On the other hand according to 
          Taylor series
          one has
          \begin{equation}
          \tilde{h}_{\mu\nu}=\sum_{j=0}^{\infty}(\tilde{h}_{\mu\nu})_{j}\Omega^j
          \label{eq:4.88}
          \end{equation}
          where $(\tilde{h}_{\mu\nu})_{0}$ represent the intrinsic metric of $\mathcal{I}$. Also  Lie derivative of the extrinsic curvature can be obtained as follows
          \begin{align}
          \label{eq:4.89}
          \mathcal{L}_{n}\tilde{K}_{\mu\nu}&=\tilde{n}^{\sigma}\underbrace{\tilde{\nabla}_{\sigma}\tilde{K}_{\mu\nu}}_{\tilde{n}^{\sigma}\tilde{\nabla}_{\sigma}\tilde{\nabla}_{\mu}\tilde{n}_{\nu}}-\underbrace{\tilde{K}_{\mu\sigma}}_{\tilde{\nabla}_{\mu}\tilde{n}_{\sigma}}\tilde{\nabla}_{\nu}\tilde{n}^{\sigma}-\underbrace{\tilde{K}_{\nu\sigma}}_{\tilde{\nabla}_{\mu}\tilde{n}_{\sigma}}\tilde{\nabla}_{\mu}\tilde{n}^{\sigma}\\
          &=\tilde{n}^{\sigma}\tilde{\nabla}_{\sigma}\tilde{\nabla}_{\mu}\tilde{n}_{\nu}-\tilde{n}^{\sigma}\tilde{\nabla}_{\mu}\tilde{\nabla}_{\sigma}\tilde{n}_{\nu}+\tilde{\nabla}_{\mu}\underbrace{(n^{\sigma}\tilde{\nabla}_{\sigma}\tilde{n}_{\nu})}_{=0}-\tilde{\nabla}_{\mu}\tilde{n}_{\sigma}\tilde{\nabla}_{\nu}\tilde{n}^{\sigma}\notag\\
          &=\tilde{n}^{\sigma}(\tilde{\nabla}_{\sigma}\tilde{\nabla}_{\mu}-\tilde{\nabla}_{\mu}\tilde{\nabla}_{\sigma})\tilde{n}_{\nu}-\tilde{\nabla}_{\mu}\tilde{n}_{\sigma}\tilde{\nabla}_{\nu}\tilde{n}^{\sigma}\notag\\
          &=\tilde{n}^{\sigma}\tilde{n}^{\rho}\tilde{R}_{\sigma\mu\rho\nu}-\tilde{K}_{\mu\sigma}\tilde{K}_{\nu}^{\sigma}.
          \end{align}
          When one multiply this relation by $\tilde{g}^{\mu\sigma}$, then
          \begin{equation}
          \mathcal{L}_{n} \tilde{K}_{\nu}^{\mu}={\mathcal{R}}_{\nu}^{\mu}+\tilde{K}\tilde{K}_{\nu}^{\mu}-\Omega^{-1}\tilde{K}\tilde{h}_{\nu}^{\mu}-4\Omega^{-1}\tilde{K}_{\nu}^{\mu}.
          \end{equation}
          
          Now for expand Einstein tensor one has (see Appendix \ref{ap:d})
          \begin{equation}
          \label{eq:4.91}
          \tilde{G}_{\mu\nu}|_{\mathcal{I}}=2 \Omega^{-1}(\tilde{K}_{\mu\nu}-\tilde{g}_{\mu\nu}\tilde{K}).
          \end{equation}
          As the conformal derivative of Einstein tensor vanishes, it is possible to write
          \begin{equation}
          \tilde{h}^{\nu}_{\mu}\tilde{G}_{\nu\sigma}n^{\sigma}=0=\tilde{h}^{\nu}_{\mu}\tilde{R}_{\nu\sigma}n^{\sigma}=D_{\nu}K^{\nu}_{\mu}-D_{\mu}K.
          \end{equation}
          Also from \eqref{eq:4.91}, we know that
          \begin{equation}
          \label{eq:4.93}
          \tilde{\mathcal{R}}+\tilde{K}^2-\tilde{K}_{\mu\nu}\tilde{K}^{\mu\nu}=4\Omega^{-1}\tilde{K}. 
          \end{equation}
          For more details see Appendix \ref{ap:d}. Also it is useful to define traceless part of the extrinsic curvature, $K_{\mu\nu}$ tensor  
          \begin{equation}
          \tilde{P}^{\mu}_{\nu}=\tilde{K}^{\mu}_{\nu}-\frac{\tilde{h}^{\mu}_{\nu}}{3}\tilde{K}.
          \end{equation}
          
          Here are other expanded parameters that would be useful for our calculations
          \begin{align}
          &\tilde{P}^{\mu}_{\nu}=\sum_{j=0}^{\infty}(\tilde{P}^{\mu}_{\nu})_{j}\Omega^j\quad,\quad \tilde{K}=\sum_{j=0}^{\infty}(\tilde{K})_{j}\Omega^j,\\
          &\tilde{R}_{\mu\nu}=\sum_{j=0}^{\infty}(\tilde{R}_{\mu\nu})_{j}\Omega^j\quad,\quad \tilde{R}=\sum_{j=0}^{\infty}(\tilde{R})_{j}\Omega^j.\notag
          \end{align}
          If Lie derivative defines as
           \begin{equation}
          \mathcal{L}_{n}\equiv\frac{d}{d\Omega},
          \end{equation}
          then
           \begin{align}
          \label{eq:4.92}
          &\frac{d}{d\Omega}\tilde{P}^{\mu}_{\nu}=[\tilde{\mathcal{R}}^{\mu}_{\nu}-\frac{\tilde{h}^{\mu}_{\nu}}{3}\tilde{\mathcal{R}}]-\tilde{K}\tilde{P}^{\mu}_{\nu}+2\Omega^{-1}\tilde{P}^{\mu}_{\nu},\\
          &\frac{d}{d\Omega}\tilde{K}=-\tilde{\mathcal{R}}-\tilde{K}^2+5\Omega^{-1}\tilde{K},\notag\\
          &\frac{d}{d\Omega}\tilde{h}^{\mu}_{\nu}=2\tilde{h}_{\nu \sigma}\tilde{K}^{\sigma}_{\mu}.\notag
          \end{align}
          Accordingly it is possible to write higher orders as \cite{jager2008conserved}
          \begin{align}
          \label{eq:4.97}
          &(2+j)(\tilde{P}^{\mu}_{\nu})_{j}=([\tilde{\mathcal{R}}^{\mu}_{\nu})_{j-1}-\frac{\tilde{h}^{\mu}_{\nu}}{3}(\tilde{\mathcal{R}})_{j-1}]-\sum_{m=0}^{j-1}(\tilde{K})_{m}(\tilde{P}^{\mu}_{\nu})_{j-1-m},\\
          &(5+j)(\tilde{K})_{j}=-(\tilde{\mathcal{R}})_{j-1}-\sum_{m=0}^{j-1}(\tilde{K})_{m}(\tilde{K})_{j-1-m},\notag\\
          &j(\tilde{h}_{\mu\nu})_{j}=2\sum^{j-1}_{m=0}[(\tilde{h}_{\nu\sigma})_{m}(\tilde{P}^{\sigma}_{\mu})_{j-1-m}+\frac{1}{3}(\tilde{h}_{\mu\nu})_{m}(\tilde{K})_{j-1-m}],\notag
          \end{align}
          which are Fefferman-Graham relations for Einstein's equations. $j$ should be smaller than $d-2$ in \eqref{eq:4.97}.
          
          To find $\tilde{h}_{\mu\nu}$ one has to write \eqref{eq:4.88} for $j=2$ as here
          $d=4$,
          \begin{equation}
          \tilde{h}_{\mu\nu}=(\tilde{h}_{\mu\nu})_0\Omega^{0}+(\tilde{h}_{\mu\nu})_1\Omega^{1}+(\tilde{h}_{\mu\nu})_2\Omega^{2}
          \end{equation}
          According to \eqref{eq:4.97} one has
          \begin{align}
          \tilde{h}_{\mu\nu}&=(\tilde{h}_{\mu\nu})_{0}+1/2 \Omega^2(\tilde{h}_{\mu\nu})_{0}+3/2\Omega^2\mathcal{R}_{\mu\nu}-3/2\Omega^2KK_{\mu\nu}+3/2\Omega K(\tilde{h}_{\mu\nu})_{0}\\
          &-6\Omega K_{\mu\nu}
          -3/2\Omega^2K^{\nu}_{\mu}K_{\nu\sigma}
          -3/2\Omega K_{\mu\sigma}\notag
          \end{align}
          On the other hand for electric part of the Weyl tensor one has
          \begin{equation}
          \label{eq:4.100}
          \tilde{E}_{\mu\nu}=\frac{1}{d-3}\Omega^{3-d}(\tilde{C}_{\mu\nu\sigma\rho}\tilde{n}^{\nu}\tilde{n}^{\rho})
          \end{equation}
          where 
          $d=4$. Someone remember the relation \eqref{eq:5.9}, on the other hand the relation for conformal transformed Schouten tensor in $\omega=constant$ surfaces is
          \begin{equation}
          \tilde{S}_{\mu\nu}=-2\Omega^{-1}\tilde{\nabla}_{\mu}\tilde{n}_{\nu}.
          \end{equation}
The Riemann tensor multiplies by $n^{\nu}n^{\rho}$ and define the other indices as free indices. Using \eqref{eq:5.9} it is possible to write
          \begin{equation}
          \label{4.102}
          \tilde{h}^{\iota}_{\mu}\tilde{h}^{\chi}_{\rho}\tilde{R}_{\iota\nu\chi\rho}n^{\nu}n^{\rho}=\tilde{h}^{\iota}_{\mu}\tilde{h}^{\chi}_{\rho}\tilde{C}_{\iota\nu\chi\rho}n^{\nu}n^{\rho}+\Omega^{-1}{K}_{\mu\sigma}
          \end{equation}
          where \eqref{eq:4.39} with $l=1$ has been used.
          
         Also for the Riemann tensor one can write
          \begin{equation}
         \label{4.103}
         \tilde{h}^{\iota}_{\mu}\tilde{h}^{\chi}_{\rho}\tilde{R}_{\iota\nu\chi\rho}n^{\nu}n^{\rho}= \tilde{h}^{\iota}_{\mu}\tilde{h}^{\chi}_{\rho}n^{\nu}(\tilde{\nabla}_{\iota}\tilde{\nabla}_{\nu}-\tilde{\nabla}_{\nu}\tilde{\nabla}_{\iota})n_{\chi}=\mathcal{L}_{n}K_{\mu\sigma}+K_{\mu\nu}K^{\nu}_{\sigma},
         \end{equation}
         using
          \eqref{4.102}
       and
         \eqref{4.103} one has
         \begin{equation}
         \tilde{C}_{\mu\nu\sigma\rho}{n}^{\nu}{n}^{\rho}=\mathcal{L}_{n}{K}_{\mu\sigma}+{K}_{\mu}^{\nu}{K}_{\nu\sigma}+\Omega^{-1}{K}_{\mu\sigma}
         \end{equation}
         \begin{equation}
         \tilde{C}_{\mu\nu\sigma\rho}{n}^{\nu}{n}^{\rho}=\mathcal{L}_{n}{K}_{\mu\sigma}+{K}_{\mu}^{\nu}{K}_{\nu\sigma}+\Omega^{-1}{K}_{\mu\sigma}.
         \end{equation}
         Using this relation in \eqref{eq:4.100} one has
         \begin{equation}
         \label{eq:4.106}
         \tilde{E}_{\mu\nu}=\Omega^{-1}(\mathcal{L}_{n}{K}_{\mu\sigma}+{K}_{\mu}^{\nu}{K}_{\nu\sigma}+\Omega^{-1}{K}_{\mu\sigma})
         \end{equation}
         According to these relations the relation of the unphysical metric in four dimensions is 
         \begin{equation}
         \label{eq:4.101}
         \tilde{g}_{\mu\nu}=-\tilde{\nabla}_{\mu}\Omega\tilde{\nabla}_{\nu}\Omega+(1+\frac{1}{2}\Omega^2)(\tilde{h}_{\mu\nu})_{0}-\frac{3}{2}\Omega^3\tilde{E}_{\mu\nu}+O(\Omega^4)
         \end{equation}
         where $(\tilde{h}_{\mu\nu})_{0}$ is the metric of a three-sphere. Unfortunately the Killing equation for this metric is not exactly solvable.
         \section{Finding the intrinsic metric according to the tetrad formalism}
         To simplify the following calculations, de Sitter line element is written as
         \begin{equation}
         ds^{2}=-F(r)dt^2+F(r)^{-1}dr^2+r^2d\Omega^2
         \end{equation}
         where $F(r)=1-\Lambda r^2/3$. Using advance null coordinates one has
         \begin{equation}
         ds^2=-F(r)du^2-2dudr+r^2d\Omega^2
         \end{equation}
         where $u=t-r^*$, then the matrix representation of the reversed metric is
         	\begin{equation}
         g^{\mu\nu}=
         \begin{bmatrix}
         0&-1&0&0\\
         -1&-F(r)&0&0\\
         0&0&r^{-2}&0\\
         0&0&0&r^{-2}\csc^2\theta
         \end{bmatrix}
         \end{equation}
         Accordingly a null tetrad can be defined \cite{lopez2006absorption,saw2016mass,saw2017behavior}
         \begin{align}
         \label{eq:4.11111}
         & {l}^{\mu}=[0,1,0,0],\\
         & {n}^{\mu}=[1,F(r),0,0]\notag,\\
         & {m}^{\mu}=[0,0,\frac{1}{\sqrt{2}r},\frac{i}{\sqrt{2}r}\csc^2\theta],\notag\\
         &{\bar{m}}^{\mu}=[0,0,\frac{1}{\sqrt{2}r},-\frac{i}{\sqrt{2}r}\csc^2\theta].\notag
         \end{align}
         Where the metric has the relation 	$g^{\mu\nu}=l^{\mu}n^{\nu}+l^{\nu}n^{\mu}-m^{\mu}\bar{m}^{\nu}-m^{\nu}\bar{m}^{\mu}$. For $r$ independent terms in ${{m}}^{\mu}$
         and  ${\bar{m}}^{\mu}$ one has
         	\begin{align}
         &\xi^{\varphi(0)}=\frac{i}{\sqrt{2}}\csc^2\theta,\\
         &\xi^{\theta(0)}=\frac{i}{\sqrt{2}}\notag
         \end{align}
         and for higher order
         \begin{align}
         \xi^{\mu}=\xi^{\mu(0)}r^{-1}+O(r^{-2}).
     	\end{align}
     	Derivative operators can be defined according to 	\eqref{eq:4.11111} \cite{saw2016mass}
     		\begin{align}
     	&D=l^{\mu}\nabla_{\mu}=\frac{\partial}{\partial r},\\
     	&D'=l^{\mu}\nabla_{\mu}=\frac{\partial}{\partial u}-1/2(1-\Lambda r^2/3)\frac{\partial}{\partial r},\notag\\
     	&\delta=m^{\mu}\nabla_{\mu}=\frac{1}{\sqrt{2}r}\frac{\partial}{\partial_{\theta}}+\frac{i}{\sqrt{2}r}\csc^2\theta\frac{\partial}{\partial_{\varphi}},\notag\\
     	&\delta'=\bar{m}^{\mu}\nabla_{\mu}=\frac{1}{\sqrt{2}r}\frac{\partial}{\partial_{\theta}}-\frac{i}{\sqrt{2}r}\csc^2\theta\frac{\partial}{\partial_{\varphi}}
     	\end{align}
         and their operations on 
         $u,r,\theta,\varphi$
         are
         \begin{align}
         &Du=0\quad,\quad D'u=1\quad,\quad \delta u=0\quad,\quad \delta' u=0,\\
         &Dr=1\quad,\quad D'r=\Lambda/6r^2-1/2\quad,\quad \delta r=0\quad,\quad \delta' r=0\notag,\\
         &D\theta=0\quad,\quad D'\theta=0\quad,\quad \delta\theta=1/\sqrt{2}r\quad,\quad \delta'\theta=1/\sqrt{2}r,\notag\\
         &D\varphi=0\quad,\quad D'\varphi=0\quad, \quad \delta\varphi=i/\sqrt{2}r\csc \theta\quad, \quad \delta'\varphi=-i/\sqrt{2}r\csc \theta.\notag
         \end{align}
         Also for second order derivatives one has
         	\begin{align}
         \label{eq:4.117}
         &DD'r=\Lambda r/3\quad,\quad D\delta\theta=-1/\sqrt{2}r^2,\\
         &D\delta\varphi=-i\csc\theta/\sqrt{2}r^2\quad,\quad D'\delta\theta=\Lambda/6\sqrt{2}+1/2\sqrt{2}r^2\notag,\\
         &D'\delta\varphi=(\Lambda/6\sqrt{2}+1/2\sqrt{2}r^2)i\csc\theta\quad,\quad \delta'\delta\varphi(-i/2r^2)=\csc\theta\cot\theta\notag,\\
         &\delta\delta'\varphi=(i/2r^2)\csc\theta\cot\theta.\notag
         \end{align}
         As the calculations have been done for $O(r^{-2})$ only terms with 	$\Lambda$ become important. So $\xi^{\mu}$ can be rewrite as follows
         \begin{align}
         &\xi^{\theta(0)}=\frac{1}{\sqrt{2}}e^{\Lambda f(u,\theta)},\\
         &\xi^{\varphi(0)}=\frac{1}{\sqrt{2}}e^{\Lambda f(u,\theta)}\csc\theta\notag.
         \end{align}
         
         Hence for spherical expanded metric one has
         \begin{equation}
         	\label{eq:11}
         g_1=e^{\Lambda f(u,\theta)}d\theta^2+e^{\Lambda f(u,\theta)}\csc\theta d\varphi^2.
         \end{equation}
         Also for first relation in \eqref{eq:4.117}, one has
         \begin{equation}
         g=\frac{\Lambda}{3}r^2du^2+r^2g_{1}+O(r^{-2}).
         \end{equation}
         Considering the conformal factor 
         $	\Omega=r^{-1}$ 
         one can obtain
         \begin{equation}
         \label{eq:13}
         \tilde{g}=\frac{\Lambda}{3}du^2+g_{1}+O(r^{-1}).
         \end{equation}
         \subsection{Killing vector fields for the intrinsic metric}
         Unfortunately it is not possible to solve Killing equations analytically   for the metric \eqref{eq:13}. On the contrary, one can write Killing equation for spherically symmetric part of the metric, that is shown in  relation \eqref{eq:11}. First 
         one  must take an arbitrary killing field $X^{\mu}=X^{\theta}\partial_{\theta}+X^{\varphi}\partial_{\varphi}$ and then write the killing equation for it \cite{saw2017mass}
         \begin{equation}
         \mathcal{L}_{X}g_{\mu\nu}=X^{\sigma}\partial_{\sigma}g_{\mu\nu}+g_{\sigma\nu}\partial_{\mu}X^{\sigma}+g_{\mu\sigma}\partial_{\nu}X^{\sigma}.
         \end{equation}
         This relation gives three independent equations
         \begin{align}
         &\theta\theta:\partial_{\theta}(X^{\theta}e^{\Lambda e^{\Lambda f}})=0,\\
         &\varphi\varphi : \partial_{\varphi}X^{\varphi}+X^{\theta}\cot\theta=0,\notag\\
         & \theta\varphi: \partial_{\varphi}X^{\theta}+\partial_{\theta}X^{\varphi}e^{-4\Lambda f}\sin^2\theta=0.\notag
         \end{align}
         That result
         \begin{align}
         \label{eq:15}
         &X^{\theta}(\theta,\varphi)=\frac{dA(\varphi)}{d\varphi}e^{-\Lambda f(\theta)},\\
         & X^{\varphi}(\theta,\varphi)=2\sqrt{2}\alpha(\theta)A(\varphi)-X(\theta),\notag\\
         &\frac{d^2A(\varphi)}{d\varphi^2}+(2\sqrt{2}e^{-3\Lambda f(\theta)}\frac{d\alpha (\theta)}{d\theta}\sin^2\theta)A{\varphi}=e^{-3\Lambda f(\theta)}\frac{dX(\theta)}{d \theta}\sin^2\theta. \notag
         \end{align}
         	Where $\alpha(\theta)=-\frac{1}{2\sqrt{2}\sin\theta}\frac{d}{d\theta}(e^{-\Lambda f(\theta)}\sin\theta)$ that is spin coefficient. Let  $\omega(\theta)^2=2\sqrt{2}e^{-3\Lambda f(\theta)}\frac{d\alpha (\theta)}{d\theta}\sin^2\theta$. The last equation in \eqref{eq:15} determines harmonic oscillator with frequency $\omega(\theta)$. The function $A(\varphi)$ is then dependent of $\theta$ and $\varphi$. Since $A(\varphi)$  must be independent of $\theta$ the only possibility is to set $\omega(\theta)$ constant and $X^{\theta}$ must be zero as well.
         Therefore, we have just one single Killing vector on this axisymmetric topological 2-sphere
         \begin{equation}
         X^{\mu}=\partial_{\varphi}
         \end{equation}
         Then one does not have the whole So(3) group on $\mathcal{I}$. 
         
	\begin{appendices}
		\chapter{Compactification\label{app:A}}
\begin{figure}
	\centering
	 \begin{tikzpicture}[scale=3]
	\filldraw[fill=green!50,draw=black!50!] (0,0mm) -- (1.73mm,1mm) arc (0:83:2mm) -- cycle ;
	 \node  at (0.15, 0.35)[green]   (a) {$\theta$}; 
	 	\filldraw[fill=red!40,draw=black!50!] (0,-10mm) -- (1mm,-8.27mm) arc (0:62:2mm) -- cycle ;
	 \node  at (0.1, -0.6)[red]   (b) {$\frac{\theta}{2}$}; 
	\draw[->] (-1.25,0) -- (1.25,0) coordinate (x axis) node[left, above]{X};
	\draw[->] (0,-1.25) -- (0,1.25) coordinate (y axis)node[left]{Y};
	\draw (0,0) circle (1cm);
	\draw[very thick,dashed,purple] (0,0) -- (-90:1cm ) ;
	\draw [->] [very thick,blue] (0,0) -- (30:1cm) ;
	\node at (0.4,0.3)[blue] {$\vec{r}$};
	\draw [very thick,dashed,orange] (0,-1) -- (30:1cm);
	\node at (0.95,0.5)[orange]{$P$};

	\end{tikzpicture}
	\caption{This illustration show how $\cot\theta/2$ can be used to shows the position of $P$. }
		\label{fig:A.2}
\end{figure}
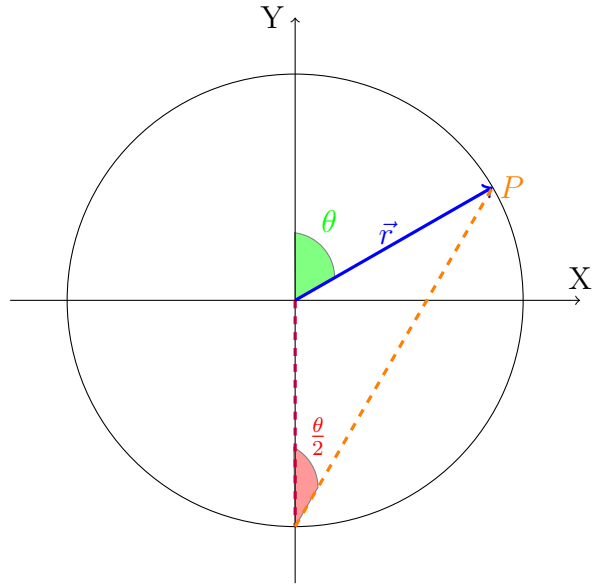

		In This appendix one can see how a manifold maps to a compactified manifold. As an example consider the map of $M=\mathbb{R}^2$ to the $\tilde{M}=\mathbb{S}^2 \backslash \{(0,0,1)\}$ where $(0,0,1)$ is the north pole of the sphere. It is possible, as figure \ref{fig:A.1} shows, to map a set of  points on the surface onto the sphere.  Considering the north pole of the sphere, N, as a fixed point and choosing a point from the surface, A, one can draw a line according to these two points. Obviously this line passes through a point on the sphere, $B$. Likewise $A$ maps onto $B$, the only point of the sphere to which no point on the surface is attributed is the North Pole. It is possible to draw an infinite number of lines which pass through the point $(0,0,1)$  that are parallel to the surface and do not cross the surface. On the other hand how farther is the point on the surface the angle between the line and the surface becomes smaller. Hence it seems to be a good idea to consider $(0,0,1)$ as the point that represents all the points on infinity of the surface $\mathbb{R}^2$. If one foliates this sphere to circles each shit has the metric
			\begin{equation}
		ds^2=dx^2+dy^2,
		\end{equation}
		with the relation
			$x^{2}+y^{2}=r^2$.
			It is possible to write the metric in the form
				\begin{equation}
			ds^2=d\xi d\xi^{*},
			\end{equation}
			where 	$\xi=x+iy$
			and
			$\xi^{*}=x-iy$. This will be helpful to change the coordinate system origin as
			$x'=x, y'=y-r, z'=z$. Thus as one can see in \ref{fig:A.2} $\cot\theta/2$ can be used to show the position of points on the circle. So  $Z=\cot\theta/2$ can be defined. In this coordinate system the metric takes the form
		\begin{equation}
	ds^2=\frac{4}{(1+Z^2)^2}dZ^2.
	\end{equation}
			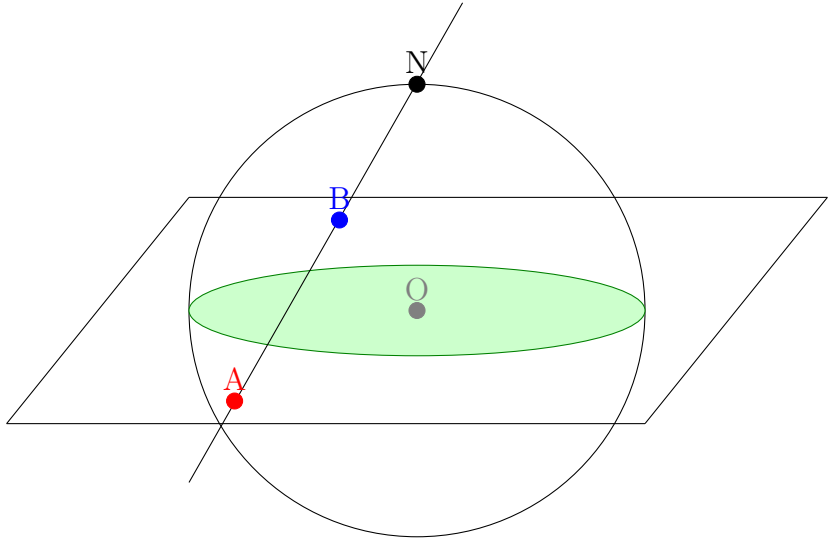
\begin{figure}
		\centering
		\begin{tikzpicture}[scale=3]
		\draw (0,0) circle (1cm);
		\filldraw[fill=green!20,draw=green!50!black](0,0) ellipse (1cm and 0.2cm);
		\draw (-1.8,-0.5) to (-1,0.5);
		\draw (-1,0.5) to (1.8,0.5);
		\draw (1.8,0.5) to (1,-0.5);
		\draw (1,-0.5) to (-1.8,-0.5);
		\draw (0.2,1.36) to (-1,-0.76) ;
		\filldraw
		(0,1) circle (1pt) node[align=center, above]{N};
		\filldraw
		(-0.8,-0.4) circle (1pt) [color=red]  node[align=center, above]{A};
		\filldraw 
		(-0.34,0.4) circle(1pt) [color=blue] node [ align=center, above]{B};
		\filldraw
		(0,0)circle(1pt) [color=gray] node [align=center, above]{O};
		\end{tikzpicture}
		\caption{Mapping $\mathbb{R}^2$ onto 	$\mathbb{S}^2$.}
		\label{fig:A.1}	
	\end{figure}
	
	It is possible to find  similar coordinates for an sphere too. First one has to set 
	$z'=z-r$,
	$y'=y$
and 
	$x'=x$. Then the position of points on the sphere can be shown by $\xi=e^{i\varphi}cot{\theta/2}$ (figure \ref{fig:A.3}). The metric can be written as
		\begin{equation}
	ds^2=d\xi d\xi^*=1/4(1+\xi \xi^*)^2(d\theta^2+\sin^2\theta d\varphi^2).
	\end{equation}
	By choosing the conformal factor $\Omega=\frac{2}{(1+\xi \xi^*)}$ the unphysical metric becomes a 2-sphere.
	
	Now consider the metric of $\mathbb{R}^{d+1}$
		\begin{equation}\label{eq:A,1}
	ds^2=-dt^2+dr^2+r^2d\Omega^2_{d-1},
	\end{equation}
	where $d\Omega^2_{d-1}$ is the metric of the 2-sphere. It is useful to define null coordinates as
	 \begin{eqnarray}
	v=t+r\quad,\quad u=t-r.
	\end{eqnarray}
	Thus the metric \eqref{eq:A,1} takes the form
		\begin{equation}
	ds^2=-dudv+\frac{1}{4}(v-u)^2 d\Omega^2_{d-1}.
	\end{equation}
	One can foliates the space to $u=constant$ or $v=constant$ surfaces and multiply the conformal factor $\Omega=\frac{2}{v-u}$ to each of these cuts. This exactly the same that one can do to find the metric near 	$\mathcal{I}$ for asymptotically flat space-times.

\tdplotsetmaincoords{60}{110}
\pgfmathsetmacro{\rvec}{.8}
\pgfmathsetmacro{\thetavec}{30}
\pgfmathsetmacro{\phivec}{60}
	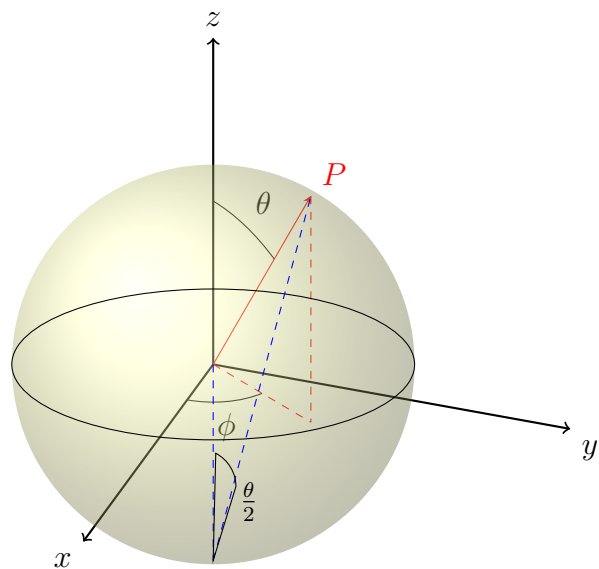
\begin{figure}
		\centering
		\begin{tikzpicture}[scale=5,tdplot_main_coords]
		\coordinate (O) at (0,0,0);
		\draw[thick,->] (0,0,0) -- (1,0,0) node[anchor=north east]{$x$};
		\draw[thick,->] (0,0,0) -- (0,1,0) node[anchor=north west]{$y$};
		\draw[thick,->] (0,0,0) -- (0,0,1) node[anchor=south]{$z$};
		\tdplotsetcoord{P}{\rvec}{\thetavec}{\phivec}
		\draw[-stealth,color=red] (O) -- (P) node[above right] {$P$} ;
		\draw[dashed, color=red] (O) -- (Pxy);
		\draw[dashed, color=red] (P) -- (Pxy);
		\tdplotdrawarc{(O)}{0.2}{0}{\phivec}{anchor=north}{$\phi$}
		\tdplotsetthetaplanecoords{\phivec}
		\tdplotdrawarc[tdplot_rotated_coords]{(0,0,0)}{0.5}{0}%
		{\thetavec}{anchor=south west}{$\theta$}
		\shade[ball color = yellow!40, opacity = 0.4] (0,0) circle (0.53cm);
		\draw (0,0)  ellipse (0.53cm and 0.2cm);
		\draw [dashed, color=blue] (0,0,0) to (0,0,-0.6);
		\draw [dashed, color=blue]   (0,0,-0.6) to (P);
		\draw[draw=black] (0,0,-0.6) -- (0.1,0.1,-0.3) arc (0:62:1mm) -- cycle ;
		\node at (0,0.1,-0.4) {$\frac{\theta}{2}$};
		\end{tikzpicture}
		\caption{ $\xi=e^{i\varphi}cot{\theta/2}$ can be used to show the position of $P$. }
		\label{fig:A.3}
	\end{figure}
	
	\chapter{Connection and Ricci tensor for higher orders}
For the metric 
\begin{equation}
g_{\mu\nu}=\hat{g}_{\mu\nu}+h_{\mu\nu},
\end{equation}
the connection can be find as 
		 \begin{align}
\label{eq:B.2}
\Gamma^{\rho}_{\mu \nu}=1/2 \hat{g}^{\rho \lambda}
(\partial_{\mu}\hat{g}_{\nu \lambda}+\partial_{\nu}\hat{g}_{\mu \lambda}-\partial_{\lambda}\hat{g}_{\mu \nu})\\
+1/2 {h}^{\rho \lambda}(\partial_{\mu}\hat{g}_{\nu \lambda}
+\partial_{\nu}\hat{g}_{\mu \lambda}-\partial_{\lambda}\hat{g}_{\mu \nu})\notag\\
+1/2 {g}^{\rho \lambda}(\partial_{\mu}{h}_{\nu \lambda}+\partial_{\nu}{h}_{\mu \lambda}-\partial_{\lambda}{h}_{\mu \nu})\notag\\
+1/2 {h}^{\rho \lambda}(\partial_{\mu}{h}_{\nu \lambda}+\partial_{\nu}{h}_{\mu \lambda}-\partial_{\lambda}{h}_{\mu \nu}).\notag
\end{align}
Using the $\hat{g}_{\sigma \epsilon}\hat{g}^{\sigma \epsilon}$ in the second term  and the definition for the zero and second order connection, 	$	\Gamma^{\rho (0)}_{\mu \nu}=1/2 \hat{g}^{\rho \lambda}(\partial_{\mu}\hat{g}_{\nu \lambda}+\partial_{\nu}\hat{g}_{\mu \lambda}-\partial_{\lambda}\hat{g}_{\mu \nu})$,
$	\Gamma^{\rho(2)}_{\mu \nu}=1/2 {h}^{\rho \lambda}(\partial_{\mu}{h}_{\nu \lambda}+\partial_{\nu}{h}_{\mu \lambda}-\partial_{\lambda}{h}_{\mu \nu})$,
the relation 	\eqref{eq:B.2} can be written as
 \begin{align}
\label{eq:B.3}
\Gamma^{\rho}_{\mu \nu}&=\Gamma^{\rho(0)}_{\mu \nu}
+1/2 {h}^{\rho \lambda}\hat{g}_{ \sigma \epsilon}\hat{g}^{ \sigma \epsilon}(\partial_{\mu}\hat{g}_{\nu \lambda}+\partial_{\nu}\hat{g}_{\mu \lambda}-\partial_{\lambda}\hat{g}_{\mu \nu})\\
& +1/2(\partial_{\mu}h^{\rho}_{\nu}+\partial_{\nu}h^{\rho}_{\mu}-\partial^{\rho}h_{\mu \nu})+\Gamma^{\rho(2)}_{\mu \nu},\notag\\
&\Rightarrow\notag\\
&\Gamma^{\rho}_{\mu \nu}=\Gamma^{\rho(0)}_{\mu \nu}
+\delta^{\lambda}_{\epsilon}h^{\rho}_{\sigma} \hat{g}^{\sigma \epsilon}(\partial_{\mu}\hat{g}_{\nu \lambda}+\partial_{\nu}\hat{g}_{\mu \lambda}-\partial_{\lambda}\hat{g}_{\mu \nu})\notag\\
&+1/2(\partial_{\mu}h^{\rho}_{\nu}+\partial_{\nu}h^{\rho}_{\mu}-\partial^{\rho}h_{\mu \nu})+\Gamma^{\rho(2)}_{\mu \nu},\notag\\
&\Rightarrow\notag\\
&\Gamma^{\rho}_{\mu \nu}=\Gamma^{\rho(0)}_{\mu \nu}+1/2(2h^{\rho}_{\sigma}\Gamma^{\rho(0)}_{\mu \nu}+\partial_{\mu}h^{\rho}_{\nu}+\partial_{\nu}h^{\rho}_{\mu}-\partial^{\rho}h_{\mu \nu})+\Gamma^{\rho(2)}_{\mu \nu}.\notag
\end{align}
On the other hand 
\begin{align}
\label{eq:B.4}
&\hat{g}^{\rho \xi}(\partial_{\mu}h_{\xi \nu}+\partial_{\nu}h_{\xi \mu}-\partial_{\xi}h_{\mu \nu}+\Gamma^{\chi}_{\mu \xi}h_{\chi \nu}+\Gamma^{\chi}_{\mu \nu}h_{\chi \xi}+\Gamma^{\chi}_{\nu \mu}h_{\chi \xi}-\Gamma^{\chi}_{\nu \xi}h_{\mu \chi}-\Gamma^{\chi}_{\mu \xi}h_{\nu\chi})\\
&=\hat{g}^{\rho \xi}(\nabla_{\mu}h_{\xi \nu}+\nabla_{\nu}h_{\xi \mu}-\nabla_{\xi}h_{\mu \nu}).\notag
\end{align}
Comparing the second term of the relation \eqref{eq:B.3} with the relation 	\eqref{eq:B.4} one can see 
	\begin{equation}
\boxed{\hat{\Gamma}^{\rho (1)}_{\mu \nu}=1/2(\hat{\nabla}_{\mu}h^{\rho}_{\nu}+\hat{\nabla}_{\nu}h^{\rho}_{\mu}-\hat{\nabla}^{\rho}h_{\mu\nu})}
\end{equation} 

Now we will calculate the Ricci tensor for higher orders
\begin{align}
R_{\mu\nu}^{(1)}&=1/2 \partial_{\rho}(\hat{\nabla}_{\mu}h^{\rho}_{\nu}+\hat{\nabla}_{\nu}h^{\rho}_{\mu}-\hat{\nabla}^{\rho}h_{\mu\nu})-1/2\partial_{\mu}(\hat{\nabla}_{\rho}h^{\rho}_{\nu}+\hat{\nabla}_{\nu}h^{\rho}_{\rho}-\hat{\nabla}^{\rho}h_{\rho\nu})\\
&+1/2(\hat{\nabla}_{\rho}h^{\rho}_{\lambda}+\hat{\nabla}_{\lambda}h^{\rho}_{\rho}-\hat{\nabla}^{\rho}h_{\rho\lambda})\Gamma^{\lambda(0)}_{\mu\nu}+1/2 (\hat{\nabla}_{\mu}h^{\lambda}_{\nu}+\hat{\nabla}_{\nu}h^{\lambda}_{\mu}-\hat{\nabla}^{\lambda}h_{\mu\nu})\Gamma^{\rho(0)}_{\rho\lambda}\notag\\
&-1/2(\hat{\nabla}_{\rho}h^{\lambda}_{\nu}+\hat{\nabla}_{\nu}h^{\lambda}_{\rho}-\hat{\nabla}^{\lambda}h_{\rho\nu})\Gamma^{\rho(0)}_{\lambda\mu}-1/2(\hat{\nabla}_{\rho}h^{\lambda}_{\mu}+\hat{\nabla}_{\mu}h^{\lambda}_{\rho}-\hat{\nabla}^{\lambda}h_{\rho\mu})\Gamma^{\rho(0)}_{\lambda\nu}\notag\\
&=1/2[\underbrace{\partial_{\rho}\hat{\nabla}_{\mu}h^{\rho}_{\nu}+\hat{\nabla}_{\mu}h^{\lambda}_{\nu}\Gamma^{\rho(0)}_{\rho\lambda}-\hat{\nabla}_{\iota}h^{\rho}_{\nu}\Gamma^{\iota(0)}_{\rho\mu}-\hat{\nabla}_{\mu}h^{\rho}_{\iota}\Gamma^{\iota(0)}_{\rho\nu}}_{\hat{\nabla}_{\rho}\hat{\nabla}_{\nu}h^{\rho}_{\mu}}]\notag\\
&+1/2[\underbrace{\partial_{\rho}\hat{\nabla}_{\nu}h^{\rho}_{\mu}+\hat{\nabla}_{\nu}h^{\lambda}_{\mu}\Gamma^{\rho(0)}_{\rho\lambda}-\hat{\nabla}_{\iota}h^{\rho}_{\mu}\Gamma^{\iota(0)}_{\rho\nu}-\hat{\nabla}_{\nu}h^{\rho}_{\iota}\Gamma^{\iota(0)}_{\rho\mu}}_{\hat{\nabla}_{\rho}\hat{\nabla}_{\mu}h^{\rho}_{\nu}}].\notag
\end{align}
Finely the following relation is obtained
	\begin{equation}
\boxed{R^{(1)}_{\mu\nu}=1/2(\hat{\nabla}_{\rho}\hat{\nabla}_{\nu}h^{\rho}_{\mu}+\hat{\nabla}_{\rho}\hat{\nabla}_{\mu}h^{\rho}_{\nu}-\hat{\nabla}^{\rho}\hat{\nabla}_{\rho}h_{\mu\nu}-\hat{\nabla}_{\mu}\hat{\nabla}_{\nu}h)}.
\end{equation}
For the second order connection
\begin{align}
\label{eq:B.8}
R^{(2)}_{\mu \nu}&= 1/2(\partial_{\rho}h^{\rho \lambda}\partial_{\mu}h_{\lambda \nu}+\partial_{\rho}h^{\rho \lambda}\partial_{\nu}h_{\mu \lambda }-\partial_{\rho}h^{\rho \lambda}\partial_{\lambda}h_{\mu \nu})\\
&+1/2h^{\rho \lambda}(\partial_{\rho}\partial_{\mu}h_{\lambda \nu}+\partial_{\rho}\partial_{\nu}h_{\mu \lambda}-\partial_{\rho}\partial_{\lambda}h_{\mu \nu})\notag\\
&-1/2 (\partial_{\nu}h^{\rho \lambda}\partial_{\mu}h_{\lambda \rho}+\partial_{\nu}h^{\rho \lambda} \partial_{\rho}h_{\mu \lambda}-\partial_{\nu}h^{\rho \lambda}\partial_{\lambda}h_{\rho \mu})\notag\\
&-1/2h^{\rho \lambda}(\partial_{\nu}\partial_{\mu}h_{\lambda \rho}+\partial_{\nu} \partial_{\rho}h_{\mu \lambda}-\partial_{\nu}\partial_{\lambda}h_{\rho \mu})\notag\\
&+1/4(\hat{\nabla}_{\rho}h^{\rho}_{\lambda}+\hat{\nabla}_{\lambda}h^{\rho}_{\rho}-\hat{\nabla}^{\rho}h_{\nu \mu})
\times(\hat{\nabla}_{\mu}h^{\lambda}_{\nu}+\hat{\nabla}_{\nu}h^{\lambda}_{\mu}-\hat{\nabla}^{\lambda}h_{\mu \nu})\notag\\
&-1/4(\hat{\nabla}_{\nu}h^{\rho}_{\lambda}+\hat{\nabla}_{\lambda}h^{\rho}_{\nu}-\hat{\nabla}^{\rho}h_{\lambda \nu})
\times(\hat{\nabla}_{\rho}h^{\lambda}_{\mu}+\hat{\nabla}_{\mu}h^{\lambda}_{\rho}-\hat{\nabla}^{\lambda}h_{\rho \mu})\notag\\
&+1/4(\hat{g}^{\rho\xi}(\partial_{\rho}\hat{g}_{\lambda \xi}+\partial_{\lambda}\hat{g}_{\xi \rho}-\partial_{\xi}\hat{g}_{\rho \lambda}))
\times h^{\lambda \xi}(\partial_{\mu}h_{\nu \xi}+\partial_{\nu}h_{\xi \mu}-\partial_{\xi}h_{\mu \nu})\notag
\\
&+1/4(h^{\rho\xi}(\partial_{\rho}h_{\lambda \xi}+\partial_{\lambda}h_{\xi \rho}-\partial_{\xi}h_{\rho \lambda}))
\times \hat{g}^{\lambda \xi}(\partial_{\mu}\hat{g}_{\nu \xi}+\partial_{\nu}\hat{g}_{\xi \mu}-\partial_{\xi}\hat{g}_{\mu \nu})\notag\\
&+1/4 \hat{g}^{\rho \xi}(\partial_{\nu}\hat{g}_{\lambda \xi}+\partial_{\lambda}\hat{g}_{\xi \nu}-\partial_{\xi}\hat{g}_{\nu \lambda})
\times
h^{\lambda \xi}(\partial_{\rho}h_{\mu \xi}+\partial_{\mu}h_{\xi \rho}-\partial_{\xi}h_{\rho \mu})\notag
\\
&+1/4h^{\rho \xi}(\partial_{\nu}h_{\lambda \xi}+\partial_{\lambda}h_{\xi \nu}-\partial_{\xi}h_{\nu \lambda})
\times
\hat{g}^{\lambda \xi}(\partial_{\rho}\hat{g}_{\mu \xi}+\partial_{\mu}\hat{g}_{\xi \rho}-\partial_{\xi}\hat{g}_{\rho \mu}).\notag
\end{align}	
In this relation many terms may be vanish depend on the gauge that we choose.
\chapter{Ricci decomposition\label{ap:j}}
It is possible to write Riemann tensor to three irreducible terms
\begin{equation}
R_{\mu \nu \sigma \rho}=C_{\mu \nu \sigma \rho}+E_{\mu \nu \sigma \rho}+G_{\mu \nu \sigma \rho}.
\label{eq:2.20}
\end{equation}
where $C_{\mu \nu \sigma \rho}$ is Weyl tensor, $E_{\mu \nu \sigma \rho}$ is semi trace-less part, $G_{\mu \nu \sigma \rho}$ is scalar part of the  Riemann  tensor. It is possible to obtain this terms using the Kulkarni-Nomizu  product. Then for $E_{\mu \nu \sigma \rho}$ one has
\begin{equation}
\label{eq:G.2}
E_{\mu \nu \sigma \rho}=\alpha (g\KN R)_{\mu \nu \sigma \rho}=2\alpha
(g_{\rho[\mu}R_{\nu]\sigma}-g_{\sigma[\mu}R_{\nu]\rho}),
\end{equation}
 Also for 	$G_{\mu \nu \sigma \rho}$ it is possible to write
 	\begin{equation}
 \label{eq:G.3}
 G_{\mu \nu \sigma \rho}=\beta R(g\KN g)_{\mu \nu \sigma \rho}	=\beta R(g_{\mu \rho}g_{\nu \sigma}-g_{\mu \sigma}g_{\nu \rho}).
 \end{equation}
 	$\alpha$
and
 $\beta$
in relations
 \eqref{eq:G.2}
 and
 \eqref{eq:G.3}
 are arbitrary constants.
 
 Putting
 \eqref{eq:G.2}
and
 \eqref{eq:G.3}
in
 \eqref{eq:2.20}, one can write
 	\begin{equation}
 \label{eq:G.4}
 R_{\mu \nu \sigma \rho}=2\beta g_{\rho[\mu}g_{\nu]\sigma}+2\alpha (g_{\rho[\mu}R_{\nu]\sigma}-g_{\sigma[\mu}R_{\nu]\rho})+C_{\mu \nu \rho \sigma}.
 \end{equation}
 Now  	$\alpha$
 and
 $\beta$ can be obtained
 \begin{align}
 \label{eq:G.5}
 &\alpha=\frac{1}{n-2},\\
 &\beta=\frac{-1}{(n-1)(n-2)}.\notag
 \end{align} 
 Using constants \eqref{eq:G.5}, the relation \eqref{eq:G.4} can be rewritten
 	\begin{equation}
 \label{eq:G.6}
 R_{\mu \nu \sigma \rho}=\frac{-2}{(n-1)(n-2)} g_{\rho[\mu}g_{\nu]\sigma}+\frac{2}{n-2} (g_{\rho[\mu}R_{\nu]\sigma}-g_{\sigma[\mu}R_{\nu]\rho})+C_{\mu \nu \rho \sigma}.
 \end{equation}
 This process is called Ricci decomposition.
 
 Also it is possible to rewrite these relation according to Schouten tensor 	$S_{\mu\nu}=R_{\mu\nu}-\frac{1}{6}g_{\mu\nu}R$, in four dimensions
 	\begin{equation}
 {R}_{\mu\nu\sigma\rho}={C}_{\mu\nu\sigma\rho}+g_{\mu[\sigma}{S}_{\rho]\nu}-g_{\nu[\sigma}{S}_{\rho]\mu}.
 \end{equation}
 or
 	\begin{equation}
 R^{\mu\nu}_{\sigma \rho}=C^{\mu\nu}_{\sigma \rho}-\frac{1}{3}R\delta^{\mu}_{[\sigma}\delta^{\nu}_{\rho]}+2\delta^{[\mu}_{[\sigma}R^{\nu]}_{\rho]}.
 \end{equation}
 
 It is possible to describe left dual right dual for Weyl tensor
 	 \begin{align}
 &^{\sim} C_{\mu\nu\sigma\rho}\equiv\frac{1}{2}\varepsilon_{\mu\nu\iota\chi}C^{\iota\chi}_{\:\:\:\:\sigma\rho},\\
 &C_{\mu\nu\sigma\rho}^{\sim}\equiv\frac{1}{2}\varepsilon_{\sigma\rho\iota\chi}C^{\:\:\:\:\iota\chi}_{\mu\nu}.\notag
 \end{align}
 Dual tensors satisfy the following relation
  \begin{equation}
 ^{\sim} C_{\mu\nu\sigma\rho}\equiv C_{\mu\nu\sigma\rho}^{\sim}.
 \end{equation}
 It is also useful to represent the complex conjugate form of the Weyl tensor
  \begin{equation}
 \star C_{\mu\nu\sigma\rho}\equiv C_{\mu\nu\sigma\rho}+iC_{\mu\nu\sigma\rho}^{\sim}.
 \end{equation}
 
 On the other hand for self-dual bivectors one can write
 \begin{equation}
 X_{\mu}\equiv \star X_{\mu\nu}u^{\nu}=0\quad,\quad X_{\mu}u^{\mu}=0\quad,\quad u_{\mu}u^{\nu}=-1,
 \end{equation}
 where  $u^{\mu}$ is a timelike unit vector. So For Weyl tensor one has
  \begin{equation}
 -Q_{\mu\nu}\equiv\star C_{\mu \nu\sigma \rho }u^{\nu}u^{\rho}\equiv E_{\mu\sigma}+iB_{\mu\sigma}\quad,\quad u_{\mu}u^{\mu}=-1,
 \end{equation}
 where  $E_{\mu\sigma} $
and 
 $ B_{\mu\sigma}$ are electric and magnetic parts of the Weyl tensor.
 
If near the de Sitter null infinity $\tilde{n}^{\mu}$ used as the timelike unit vector
 with the relation $\tilde{n}^{\mu}\tilde{n}_{\mu}=-1/l^2$, then one has
  \begin{align}
 &\tilde{E}_{\mu\sigma}:=l^{2}\tilde{C}_{\mu\nu\sigma\lambda}n^{\nu}n^{\sigma},\\
 &\tilde{B}_{\mu\sigma}:=l^{2} \star \tilde{C}_{\mu\nu\sigma\lambda}n^{\nu}n^{\sigma}.\notag
 \end{align}
 \eqref{eq:G.6} calculation for de Sitter space-time shows that $C_{\mu \nu \rho \sigma}$ and $E_{\mu \nu \rho \sigma}$ become zero. Thus Riemann tensor takes the form
 	\begin{equation}
 R_{\mu \nu \sigma \rho}=\frac{R}{n(n-1)}g_{\rho[\mu}g_{\nu]\sigma}=\frac{R}{n(n-1)}(g_{\mu \rho}g_{\nu \sigma}-g_{\mu \sigma}g_{\nu \rho}). 
 \label{eq:2.23}
 \end{equation}
 The relation \eqref{eq:2.23} shows that Riemman tensor can be described  only with the Ricci scalar that means having the curvature of a point one is able to obtain the curvature of the whole maximally symmetric space-time
 	\begin{equation}
 G_{\mu \nu}=R_{\mu \nu}-1/2Rg_{\mu \nu}=-1/4Rg_{\mu \nu}.
 \end{equation}
 
 Another traceless tensor is the Bach tensor that is also invariant under conformal transformations. Bach tensor has the following form on $\mathcal{I}$
 	\begin{equation}
 B_{\mu\nu\sigma}=D_{[\mu}(\mathcal{R}_{\nu]\sigma}-\frac{1}{4}q_{\nu]\sigma}\mathcal{R}),
 \end{equation}
 where $q_{\nu\sigma}$ is the metric on $\mathcal{I}$ and $D_{\mu}$ is the covariant derivative related to $q_{\nu\sigma}$.
 	$\mathcal{R}$
 and
 $\mathcal{R}_{\nu\sigma}$ are Ricci scalar and Ricci tensor on $\mathcal{I}$. Now it is possible to rewrite the relation of $B_{\mu\nu\sigma}$
 according to asymptotic Weyl tensor, 	$K_{\mu\nu\sigma\rho}=\Omega^{-1}{C}_{\mu\nu\sigma\rho}=\Omega$
 	\begin{equation}
 B_{\mu\nu\sigma}=\frac{1}{2}q^{\chi}_{\mu}q_{\nu}^{\psi}q_{\sigma}^{\iota}D_{[\chi}S_{\psi]\iota}=\frac{1}{2}q^{\chi}_{\mu}q_{\nu}^{\psi}q_{\sigma}^{\iota}K_{\chi\psi\iota\gamma}\tilde{n}^{\gamma}.
 \end{equation}
 \chapter{Tensor field caused by derivative operator \label{ap:d}}
 Two derivative operator $\nabla$ and 	$\tilde{\nabla}$ according to ${g}_{\mu\nu}$ and $\tilde{g}_{\mu\nu}$. If one operate them on the dual vector field, $\omega_{\nu}$ and the scalar field, $f$, according to Leibnitz rule it is possible to write \cite{stephani2009exact}
 \begin{equation}
 \tilde{\nabla}_{\mu}(f\omega_{\nu})-{\nabla}_{\mu}(f\omega_{\nu})=f(	\tilde{\nabla}_{\mu}\omega_{\nu}-{\nabla}_{\mu}\omega_{\nu}).
 \end{equation}
 Each derivative operator maps a tensor of rank $(k,l)$ to a tensor of rank  $(k,l+1)$. So $\tilde{\nabla}_{\mu}\omega_{\nu}-{\nabla}_{\mu}\omega_{\nu}$ create a tensor field, $C^{\sigma}_{\mu\nu }$ as
 	\begin{equation}
 \tilde{\nabla}_{\mu}\omega_{\nu}={\nabla}_{\mu}\omega_{\nu}-C^{\sigma}_{\mu\nu }\omega_{\sigma}.
 \end{equation}
  considering $	\tilde{\nabla}_{\mu}$ as the partial derivative $\partial_{\mu}$ then $C^{\sigma}_{\mu\nu }$ reduce to the Christoffel symbol. 
  
  It has to be said that from torsion free condition for $C^{\sigma}_{\mu\nu }$ one has the relation 
    \begin{equation}
  C^{\sigma}_{\mu\nu }=C^{\sigma}_{\nu\mu }.
  \end{equation}
  The operation of derivative operator on the metric is then
   \begin{equation}
  \label{eq:d.4}
  0=\tilde{\nabla}_{\mu}\tilde{g}_{\nu\sigma}={\nabla}_{\mu}\tilde{g}_{\nu\sigma}-C^{\rho}_{\mu\nu}\tilde{g}_{\rho\sigma}-C^{\rho}_{\mu\sigma}\tilde{g}_{\nu\rho}
  \end{equation}
  and for other indices' combination one also has
  	\begin{align}
  \label{eq:d.5}
  \tilde{\nabla}_{\nu}\tilde{g}_{\mu\sigma}&={\nabla}_{\nu}\tilde{g}_{\mu\sigma}-C^{\rho}_{\nu\mu}\tilde{g}_{\rho\sigma}-C^{\rho}_{\nu\sigma}\tilde{g}_{\mu\rho},\\
  \tilde{\nabla}_{\sigma}\tilde{g}_{\mu\nu}&={\nabla}_{\sigma}\tilde{g}_{\mu\nu}-C^{\rho}_{\sigma\mu}\tilde{g}_{\rho\nu}-C^{\rho}_{\sigma\nu}\tilde{g}_{\nu\rho}.\notag
  \end{align}
  Subtracting the first relation of \eqref{eq:d.5} from \eqref{eq:d.4} and using the second relation of \eqref{eq:d.5} one has
  	\begin{equation}
  C^{\rho}_{\mu\sigma}=1/2\tilde{g}^{\nu\rho}({\nabla}_{\mu}\tilde{g}_{\nu\sigma}+{\nabla}_{\sigma}\tilde{g}_{\mu\nu}-{\nabla}_{\nu}\tilde{g}_{\mu\sigma}).
  \label{d.6}
  \end{equation}
  For the unphysical metric one has
  \begin{equation}
  {\nabla}_{\sigma} \tilde{g}_{\mu\nu}={\nabla}_{\sigma}(\Omega^2{g}_{\mu\nu})=2\Omega{g}_{\mu\nu}\nabla \Omega.
  \end{equation}
  So the relation \eqref{d.6} can be rewritten as
  \begin{equation}
  \label{d.8}
  {C}^{\rho}_{\mu\sigma}=\Omega^{-1}{g}^{\nu\rho}({g}_{\nu\sigma}n_{\mu}+{g}_{\mu\nu}n_{\sigma}-{g}_{\mu\sigma}n^{\rho})=2\Omega^{-1}\delta^{\rho}_{(\sigma}n_{\mu)}-\Omega^{-1}{g}^{\nu\rho}{g}_{\mu\sigma}n^{\rho},
  \end{equation}
  where $n_{\mu}={\nabla}_{\mu}\Omega$. It is possible to write similar calculation according to $\tilde{\nabla}_{\mu}$ and ${g}_{\mu\nu}$
  \begin{equation}
  \label{eq:D.99}
  \tilde{C}^{\rho}_{\mu\sigma}=\Omega^{-1}{g}^{\nu\rho}({g}_{\nu\sigma}\tilde{n}_{\mu}+{g}_{\mu\nu}\tilde{n}_{\sigma}-{g}_{\mu\sigma}\tilde{n}^{\rho})=2\Omega^{-1}\delta^{\rho}_{(\sigma}\tilde{n}_{\mu)}-\Omega^{-1}{g}^{\nu\rho}{g}_{\mu\sigma}\tilde{n}^{\rho}.
  \end{equation}
  Riemann tensor can be obtained according to each
  \eqref{eq:D.99} or \eqref{d.8} but because of our convention it is important to use $\tilde{C}^{\rho}_{\mu\sigma}$ from  \eqref{eq:D.99}, so one has
  \begin{align}
  \tilde{R}^{\rho}_{\mu\sigma\nu}&={R}^{\rho}_{\mu\sigma\nu}-2\tilde{\nabla}_{[\mu}\tilde{C}^{\rho}_{\sigma]\nu}+2\tilde{C}^{\lambda}_{\nu]\mu}\tilde{C}^{\rho}_{\sigma]\lambda}\\
  &={R}^{\rho}_{\mu\sigma\nu}+2\Omega^{-1}\delta_{[\mu}\tilde{\nabla}_{\sigma]}\tilde{\nabla}_{\nu}\Omega-2\Omega^{-1}\tilde{g}^{\rho\lambda}\tilde{g}_{\nu[\mu}\tilde{\nabla}_{\sigma]}\tilde{\nabla}_{\lambda}\Omega+2\Omega^{-2}\tilde{\nabla}_{[\mu}\Omega\delta^{\lambda}_{\sigma]}\tilde{\nabla}_{\nu}\Omega\notag\\
  &-2\Omega^{-2}\tilde{\nabla}_{[\mu}\Omega\tilde{g}_{\sigma]\nu}\tilde{\nabla}_{\xi}\Omega-2\tilde{g}_{\nu[\mu}\delta^{\rho}_{\sigma]}\tilde{g}^{\lambda\xi}\tilde{\nabla}_{\xi}\Omega\tilde{\nabla}_{\lambda}\Omega\notag
  \end{align}
  and also for the Ricci tensor
  
  \begin{align}
  \label{eq:D.8}
  \tilde{R}_{\mu\nu}=&\tilde{\nabla}_{\rho}\tilde{C}^{\rho}_{\mu\nu}-\tilde{\nabla}_{\nu}\tilde{C}^{\rho}_{\rho\mu}+\tilde{C}^{\rho}_{\rho\lambda}\tilde{C}^{\lambda}_{\mu\nu}-\tilde{C}^{\rho}_{\mu\lambda}\tilde{C}^{\lambda}_{\rho\nu}\\
  =&R_{\mu\nu}+(d-2)\Omega^{-2}\tilde{\nabla}_{\nu}\Omega\tilde{\nabla}_{\mu}\Omega-(d-2)\Omega^{-2}\tilde{g}_{\mu\nu}\tilde{g}^{\rho\sigma}\tilde{\nabla}_{\rho}\Omega\tilde{\nabla}_{\sigma}\Omega\notag\\
  -&(d-2)\Omega^{-1}\tilde{\nabla}_{\mu}\tilde{\nabla}_{\nu}\Omega-\Omega^{-1}\tilde{g}_{\mu\nu}\tilde{g}^{\rho\sigma}\tilde{\nabla}_{\rho}\tilde{\nabla}_{\sigma}\Omega\notag.
  \end{align}
  multiplying this relation by $\tilde{g}_{\mu\nu}$ the Ricci scalar can be obtained 
  \begin{equation}
  \tilde{R}=R-2(d-1)\Omega^{-1}\tilde{\nabla}^{\nu}\tilde{\nabla}_{\nu}\Omega-(d-2)(d-1)\Omega^{-2}\tilde{\nabla}^{\nu}\Omega\tilde{\nabla}_{\nu}\Omega.
  \label{eq:D.9}
  \end{equation}
  On the other hand according to \eqref{eq:D.8}
  and
  \eqref{eq:D.9} Schouten tensor in four dimensions can be written as follows
  \begin{equation}
  \tilde{S}_{\mu\nu}=\tilde{R}_{\mu\nu}-1/6\tilde{g}_{\mu\nu}\tilde{R}={S}_{\mu\nu}-2\Omega^{-2}\tilde{\nabla}_{\nu}\Omega\tilde{\nabla}_{\mu}\Omega+2\Omega^{-1}\tilde{\nabla}_{\nu}\tilde{\nabla}_{\mu}\Omega.
  \end{equation}
  Einstein tensor can be also obtained using 
  \eqref{eq:D.8}
  and
  \eqref{eq:D.9}
  \begin{align}
  \label{eq:D.14}
  \tilde{G}_{\mu\nu}&=\tilde{R}_{\mu\nu}-1/2\tilde{g}_{\mu\nu}\tilde{R}=G_{\mu\nu}+ 2 \Omega^{-1}(\tilde{\nabla}_{\mu}\tilde{n}_{\nu}-\tilde{g}_{\mu\nu}\tilde{\nabla}^{\sigma}\tilde{n}_{\sigma})+3\Omega^{-2} \tilde{g}_{\mu\nu}\tilde{n}^{\sigma}\tilde{n}_{\sigma}\\
  &+\Omega^{-2} \Lambda \tilde{g}_{\mu\nu}.\notag
  \label{eq:D.13}
  \end{align} 
  from vacuum condition
  $G_{\mu\nu}=0$, thus the two final terms in \eqref{eq:D.14} simplify together, on $\mathcal{I}$
  \begin{equation}
  \tilde{G}_{\mu\nu}=2 \Omega^{-1}(\tilde{\nabla}_{\mu}\tilde{n}_{\nu}-\tilde{g}_{\mu\nu}\tilde{\nabla}^{\sigma}\tilde{n}_{\sigma})=2 \Omega^{-1}(\tilde{K}_{\mu\nu}-\tilde{g}_{\mu\nu}\tilde{K})
  \label{eq:D.15}
  \end{equation}
  where $\tilde{K}_{\mu\nu}=\tilde{\nabla}_{\mu}\tilde{n}_{\nu}$
  and
  $\tilde{K}=\tilde{\nabla}^{\sigma}\tilde{n}_{\sigma}$.
  For Riemann tensor in three dimensions one has
   \begin{equation}
  \label{eq:d.166}
  \mathcal{R}^{\sigma}_{\mu\nu\rho}\omega_{\sigma}= D_{\mu}D_{\nu}\omega_{\rho}- D_{\nu}D_{\mu}\omega_{\rho}
  \end{equation}
  where  $ D_{\mu}D_{\nu}\omega_{\rho}$ can be written as
   \begin{align}
  \label{eq:d.177}
  D_{\mu}D_{\nu}\omega_{\rho}&=D_{\mu}(h^{\psi}_{\nu}h^{\chi}_{\rho}\tilde{\nabla}_{\psi}\omega_{\chi})=h^{\iota}_{\mu}   h^{\xi}_{\nu}h^{\epsilon}_{\rho} \tilde{\nabla}_{\iota}(h^{\psi}_{\xi}h^{\chi}_{\epsilon}\tilde{\nabla}_{\psi}\omega_{\chi})\\
  &=h^{\iota}_{\mu}   h^{\xi}_{\nu}h^{\epsilon}_{\rho} \tilde{\nabla}_{\iota}(\underbrace{h^{\psi}_{\xi}}_{\tilde{g}^{\psi}_{\xi}+\tilde{n}^{\psi}\tilde{n}_{\xi}})h^{\chi}_{\epsilon}\tilde{\nabla}_{\psi}\omega_{\chi}
  +h^{\iota}_{\mu}   h^{\xi}_{\nu}h^{\epsilon}_{\rho} h^{\psi}_{\xi}\tilde{\nabla}_{\iota}(\underbrace{h^{\chi}_{\epsilon}}_{\tilde{g}^{\chi}_{\epsilon}+\tilde{n}^{\chi}\tilde{n}_{\epsilon}})\tilde{\nabla}_{\psi}\omega_{\chi}\notag\\
  &+h^{\iota}_{\mu}   h^{\xi}_{\nu}h^{\epsilon}_{\rho} h^{\psi}_{\xi}h^{\chi}_{\epsilon}\tilde{\nabla}_{\iota}\tilde{\nabla}_{\psi}\omega_{\chi}\notag
  \end{align}
  where  $\tilde{g}_{\mu\nu}=-\tilde{n}_{\mu}\tilde{n}_{\nu}+\tilde{h}_{\mu\nu}$. If one use $\tilde{\nabla}^{\mu}\tilde{n}_{\nu}=\tilde{K}^{\mu}_{\nu}$
  and
  $\tilde{\nabla}^{\rho}\tilde{g}_{\mu\nu}$ in this relation, the relation  \eqref{eq:d.177} takes the following form
   \begin{align}
  D_{\mu}D_{\nu}\omega_{\rho}&=h^{\iota}_{\mu}   h^{\xi}_{\nu}h^{\epsilon}_{\rho} h^{\psi}_{\xi}h^{\chi}_{\epsilon}\tilde{\nabla}_{\iota}\tilde{\nabla}_{\psi}\omega_{\chi}+h_{\rho}^{\chi}\tilde{K}_{\mu\nu}\tilde{n}^{\sigma}\tilde{\nabla}_{\sigma}\omega_{\chi}+h^{\sigma}_{\nu}\tilde{K}_{\mu\rho}\tilde{n}^{\xi}\tilde{\nabla}_{\sigma}\omega_{\xi}.
  \end{align}
  remember that $\tilde{K}^{\mu}_{\nu}$ indices are lowered and raised
  with $h_{\mu\nu}$. This calculation can be done similarly for  $D_{\nu}D_{\mu}\omega_{\rho}$. Then puting these relations in \eqref{eq:d.166} one has \cite{feng2018weiss}
   \begin{equation}
  \mathcal{R}^{\sigma}_{\mu\nu\rho}=h^{\iota}_{\mu}   h^{\xi}_{\nu}h^{\epsilon}_{\rho} h^{\sigma}_{\chi}\tilde{R}_{\iota\xi\epsilon}^{\chi}-\tilde{K}_{\mu\rho}\tilde{K}^{\sigma}_{\nu}-\tilde{K}_{\sigma\rho}\tilde{K}^{\sigma}_{\mu}.
  \end{equation}
  Using this relation one can also find the three-dimensional form of the Ricci tensor for $\Omega= constant$ surfaces
   \begin{equation}
  \label{eq:d.20}
  \mathcal{R}_{\mu\rho}=\mathcal{R}^{\sigma}_{\mu\sigma\rho}=\tilde{h}^{\iota}_{\mu}\tilde{h}^{\chi}_{\rho}\tilde{R}_{\iota\chi}-\tilde{K}_{\mu\rho}\tilde{K}-\tilde{K}_{\sigma\rho}\tilde{K}^{\sigma}_{\mu}.
  \end{equation}
  If one put $G_{\mu\nu}=\tilde{R}_{\mu\nu}-1/2\tilde{g}_{\mu\nu}\tilde{R}$ in \eqref{eq:D.15} and multiplying the resulting equation by  $\tilde{n}^{\mu}\tilde{n}^{\nu}$, one finds
  \begin{align}
  \label{eq:d.16}
  \tilde{R}+2\tilde{R}_{\mu\sigma}\tilde{n}^{\mu}\tilde{n}^{\sigma}=4\Omega^{-1}(\tilde{K}_{\mu\sigma}-\tilde{g}_{\mu\sigma}\tilde{K})\tilde{n}^{\mu}\tilde{n}^{\sigma}.
  \end{align}
  
  Using 
  $K_{\mu\sigma}n^{\mu}=0$ in \eqref{eq:d.16} one gets
   \begin{align}
   \label{eq:d.22}
  \tilde{R}+2\tilde{R}_{\mu\sigma}\tilde{n}^{\mu}\tilde{n}^{\sigma}=-4\Omega^{-1}\tilde{g}_{\mu\sigma}\tilde{n}^{\mu}\tilde{n}^{\sigma}\tilde{K}=4\Omega^{-1}(-\tilde{K}+\underbrace{\tilde{h}_{\mu\sigma}\tilde{n}^{\mu}\tilde{n}^{\sigma}}_{=0})=-4\Omega^{-1}\tilde{K}.
  \end{align}
  Doing some mathematics one can find the following relation by using  \eqref{eq:d.20} in \eqref{eq:d.22} so
  \begin{align}
  \mathcal{R}+\tilde{K}^2-\tilde{K}_{\mu\nu}\tilde{K}^{\mu\nu}=4\Omega^{-1}\tilde{K}.
  \end{align}
 
	\end{appendices} 
	 
\bibliographystyle{ieeetr}
\bibliography{ref}
\end{document}